\documentclass[journal=jacsat,manuscript=article]{achemso}

\setkeys{acs}{usetitle = true}
\setkeys{acs}{usetitle = true}

\usepackage[version=3]{mhchem} 
\usepackage{amssymb}
\usepackage{graphicx}
\usepackage{dcolumn}
\setkeys{acs}{usetitle = true}
\newcommand{\cm}{cm$^{-1}$}

\author{A. Sieg}
\affiliation{Physikalisches Institut, Universit{\"a}t Freiburg, 
79104 Freiburg, Germany}
\author{J. von Vangerow}
\affiliation{Physikalisches Institut, Universit{\"a}t Freiburg, 
79104 Freiburg, Germany}
\author{F. Stienkemeier}
\affiliation{Physikalisches Institut, Universit{\"a}t Freiburg, 
79104 Freiburg, Germany}
\author{O. Dulieu}
\affiliation{Laboratoire Aim{´e} Cotton, CNRS, Universit{´e} Paris-Sud, ENS Cachan, Universit{´e} Paris-Saclay, 91405 Orsay Cedex, France}
\author{M. Mudrich}
\email{mudrich@physik.uni-freiburg.de}
\phone{+49 (0)761 2038405}
\fax{+49 (0)761 2037611}
\affiliation{Physikalisches Institut, Universit{\"a}t Freiburg, 
79104 Freiburg, Germany}

\title{Desorption Dynamics of Rb$_2$ Molecules off the Surface of Helium Nanodroplets}

\keywords{Helium nanodroplets, ion imaging, photodissociation, cluster relaxation}

\begin{document}
\setkeys{acs}{usetitle = true}


\begin{abstract}
The desorption dynamics of rubidium dimers (Rb$_2$) off the surface of helium nanodroplets induced by laser excitation is studied employing both nanosecond and femtosecond ion imaging spectroscopy. Similarly to alkali metal atoms, we find that the Rb$_2$ desorption process resembles the dissociation of a diatomic molecule. However, both angular and energy distributions of detected Rb$_2^+$ ions appear to be most crucially determined by the Rb$_2$ intramolecular degrees of freedom rather than by those of the Rb$_2$He$_N$ complex. The pump-probe dynamics of Rb$_2^+$ is found to be slower than that of Rb$^+$ pointing at a weaker effective guest-host repulsion for excited molecules than for single atoms. 
\end{abstract}

\maketitle

\section{\label{sec:Intro}Introduction}
The dynamics of pure and doped helium (He) nanodroplets induced by laser excitation is currently being studied both experimentally and theoretically by various groups~\cite{Krotscheck:2001,Barranco:2006,Hernando:2012,Mudrich:2014,Ziemkiewicz:2015}. The goal is to better understand the dynamic response of a microscopic superfluid to an impulsive perturbation. Furthermore, dopant atoms and molecules embedded in He nanodroplets exhibit various peculiar dynamical phenomena upon electronic excitation or ionization such as solvation into or desolvation out of the droplets~\cite{Stienkemeier:2006,Loginov:2007,Smolarek:2010,Theisen:2011,Zhang:2012,KautschPRA:2012,Fechner:2012,Vangerow:2014,Vangerow:2015}, complex formation of dopants with He atoms~\cite{Bruehl:2001,Loginov:2007,DoeppnerJCP:2007,Mudrich:2008,Loginov:2011,Giese:2012,Leal:2014,Vangerow:2015}, or the light-induced collapse of metastable structures formed by aggregation in droplets~\cite{Przystawik:2008,Goede:2013}. Recently, the time-resolved study of rotational wave packet motion of molecules embedded in He droplets revealed that  molecular rotation is significantly slowed down compared to isolated molecules and that rotational recurrences were completely absent~\cite{PentlehnerPRL:2013}. This is in contrast to our previous observation of long-lasting coherent vibrational motion of Na$_2$, K$_2$, Rb$_2$, and Rb$_3$ molecules formed on the surface of He droplets~\cite{Claas:2006,Claas:2007,Mudrich:2009,Gruner:2011,Giese:2011} which we interpreted in terms of weak system-bath couplings~\cite{Schlesinger:2010,Gruner:2011}. However, no clear signature of the desorption process of Rb$_2$ off the droplet surface was seen so that some uncertainty has remained with regard to the location of the molecule after excitation -- on the droplet surface or in the vacuum. 


In the present work we report on an ion imaging study of the desorption dynamics of Rb$_2$ molecules formed on the surface of He nanodroplets. This work extends previous experimental studies of the desorption process of alkali (Ak) metal atoms on the one hand~\cite{Loginov:2011,Hernando:2012,Fechner:2012,Vangerow:2014,Loginov:2014,Vangerow:2015,Loginov:2015}, and on of the vibrational wavepacket dynamics of Ak dimers and trimers formed on He droplets on the other~\cite{Claas:2006,Claas:2007,Mudrich:2009,Gruner:2011,Giese:2011}. Ak metals are particularly well-suited for these studies due to their location in shallow dimple states at the droplet surface~\cite{Dalfovo:1994,Ancilotto:1995,Stienkemeier:1995}. Upon electronic excitation, Ak metal atoms and  molecules tend to desorb off the He droplet as a consequence of repulsive interaction caused by the overlap of their extended valence orbitals with the surrounding 
He~\cite{Reho:2000,Schulz:2001,Callegari:2011}. The only known exceptions are Rb and Cs atoms excited to their lowest excited states~\cite{Auboeck:2008,Theisen:2011}. Ions of Ak metals, in contrast, tend to be submerged in the He droplet interior where they form stable AkHe$_n$ `snowball' complexes due to strong attractive polarization forces~\cite{Buzzacchi:2001,Mueller:2009,Theisen:2010}. Thus, in femtosecond (fs) pump-probe experiments, where a first pump pulse excites the Ak atom and a second probe pulse ionizes it, the competing dynamics of desorption of the excited Ak atom versus the fall-back of the atom into the droplet upon ionization can be followed in real-time~\cite{Vangerow:2015}. 

For low-lying excited states, the dynamics of the desorption process is well described by the pseudo-diatomic model which treats the whole He droplet, He$_N$, as one constituent atom of the AkHe$_N$ pseudodiatomic molecule~\cite{Stienkemeier:1996,LoginovPRL:2011,Callegari:2011,Loginov:2015}. 
This approach provides a reasonably precise assignment of the spectral features in absorption spectra of the AkHe$_N$ complex~\cite{Stienkemeier:1996,Buenermann:2007,Lackner:2011}, rationalizes kinetic energies of desorbed Ak atoms, and explains the angular distributions of desorbed Ak atoms in terms of the symmetries of ground and excited pseudo-diatomic molecular states~\cite{Loginov:2011,Fechner:2012,Hernando:2012,Loginov:2014,Vangerow:2014,Loginov:2015}. Moreover, fs pump-probe transients revealing the desorption dynamics of Ak atoms are reasonably well interpreted using the pseudo-diatomic model~\cite{Vangerow:2015}. 

For Ak dimers, trimers and larger Ak$_n$ oligomers attached to He droplets the dynamics induced by electronic excitation is not as well established as for single atoms. On the one hand, Pauli repulsion resulting from the overlapping of a diffuse excited state-electron distribution with the surrounding He tends to be weaker for large molecules and clusters because the relative change of orbital radius upon electronic excitation generally is smaller than for single atoms excited in a similar energy range. Besides, larger clusters are more stably bound to the He droplet with increasing cluster size as a result of attractive dispersion forces~\cite{Stark:2010}. Note that polyatomic molecules generally remain attached to He nanodroplets upon electronic excitation into low-lying excited states~\cite{Stienkemeier:2001,Stienkemeier:2006}. Thus, the desorption probability tends to be reduced for larger Ak metal clusters. On the other hand, the increasing density of internal states may enhance slow evaporation of Ak$_n$ following excitation due to energy redistribution from intra-cluster (Ak$_n$) to inter-cluster (Ak$_n$-He$_N$) degrees of freedom. The present work clearly shows that Rb$_2$ molecules desorb off the surface of He nanodroplets when excited into intermediate electronic states. However, the desorption dynamics is more intricate than in the case of Ak atoms and appears to be most crucially determined by the Rb$_2$ molecular states rather than by coupled states of the Rb$_2$He$_N$ complex as for He droplets doped with single Ak atoms.

\section{\label{sec:experimental}Experiment}
The experiments presented here are performed using the same setup 
as described previously~\cite{Fechner:2012,Vangerow:2014,Vangerow:2015}. In short, a continuous beam of He nanodroplets with a mean size of about 20,000 He atoms per droplet is produced by a continuous expansion of He out of a cryogenic nozzle ($T_0=15$ K) with a diameter of $5\,\mu$m~\cite{Toennies:2004,Stienkemeier:2006}. 
The resulting average droplet size is about 20,000 He atoms per droplet.
An adjacent vacuum chamber contains a vapor cell filled with bulk metallic Rb heated to around 100$^\circ$C to achieve the maximum signal of Rb$_2^+$ dimer ions while keeping larger oligomer masses below the detection limit. Under these conditions He droplets are doped with two Rb atoms which subsequently form Rb$_2$ molecules on He droplets. The released binding energy of the molecules is dissipated by evaporation of He atoms from the droplets. In this formation process by trend Ak molecules in the more loosely bound triplet metastable state remain attached to the droplets. However, for larger He droplets as we use in this study, substantial amounts of Ak molecules in their singlet groundstate and even small clusters in low-spin states are also being formed on the droplets~\cite{Stienkemeier2:1995,Claas:2006,Nagl_jcp:2008,Mudrich:2009,Theisen:2011}. 

In the detector chamber further downstream, the doped He droplet beam intersects at right angles the beam of either a tunable dye laser (Sirah Cobra, pulse length $~10$\,ns, pulse energy $~1\,\mu$J, repetition rate $1\,$kHz, focal length of the focusing lens $f=12$~cm) using Coumarine 102 and Exalite 411 or that of a fs laser (Coherent Legend, pulse length $120$\,fs, pulse energy $~20\,\mu$J, repetition rate $5\,$kHz, $f=30\,$cm for both pump and probe). Photoions are detected using a standard velocity map imaging (VMI) spectrometer~\cite{Fechner:2012}. Ion images are recorded selectively for Rb$_2^+$-ions by gating the imaging detector at the corresponding ion time-of-flight. The laser is linearly polarized along the direction of the He droplet beam, which is perpendicular to the symmetry axis of the VMI spectrometer. We record single events per image frame and determine the coordinates using the centroid method. Typically $10^5$ events are summed up
in the images recorded with the ns laser. For the fs experiment, the delay between pump and probe pulse is adjusted by a movable delay stage from 0 up to 10$\,$ps with 50$\,$fs increments. Images with $10^4$ events are recorded for each delay step. VMIs are transformed into kinetic energy distributions using the Abel inversion routine pBasex~\cite{Garcia:2004}.

\begin{figure}
	\centering
	\includegraphics[width=0.8\textwidth]{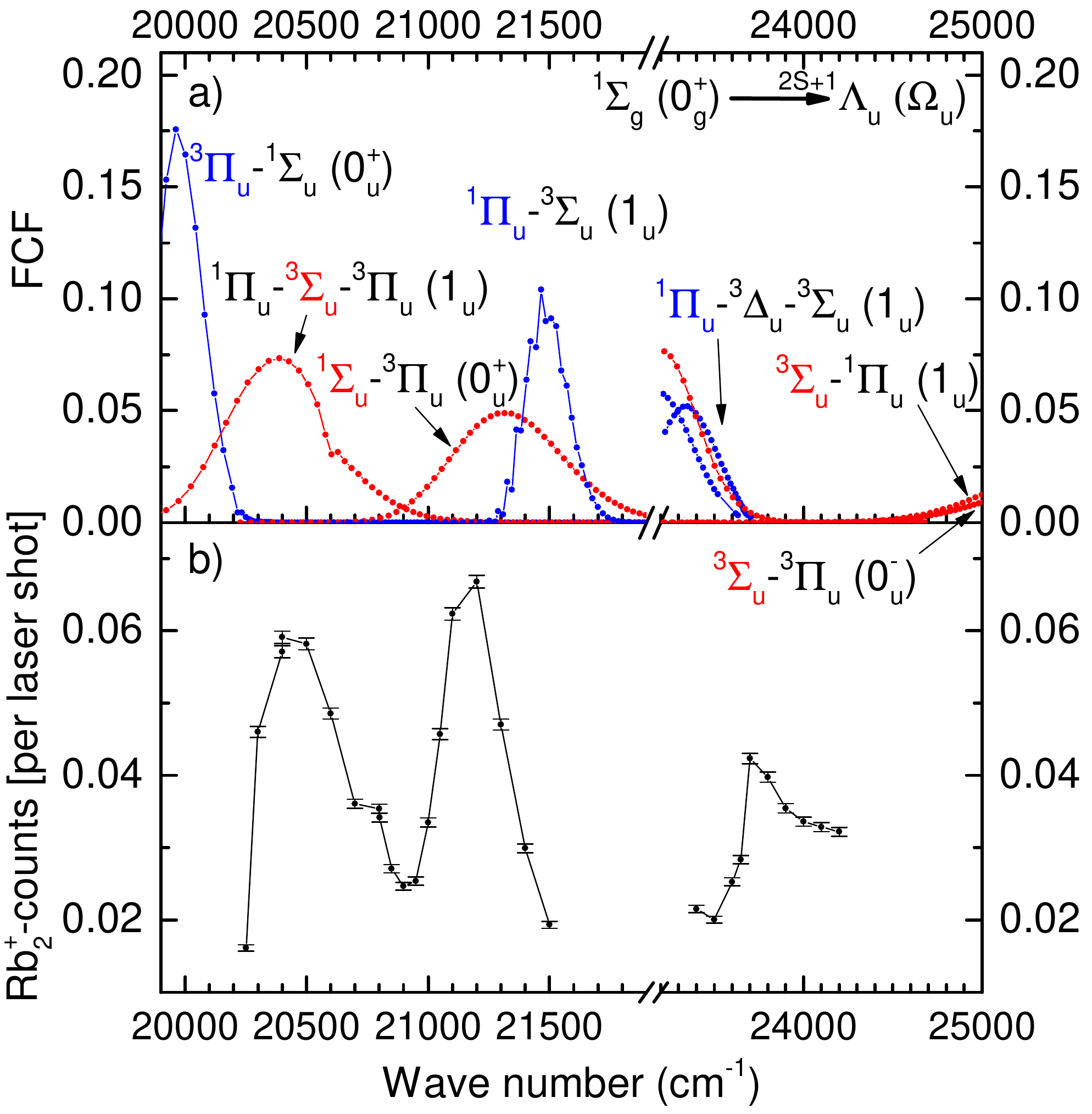}
	\caption{a) Franck-Condon-factors of allowed $g\rightarrow u$-transitions between the singlet groundstate and excited spin-orbit coupled states of Rb$_2$. The red ($\Sigma$) and blue ($\Pi$) colors indicate the main (non-spin-orbit) character of the relevant potential energy curves in the range of laser excitation. b) Measured yield of Rb$_2^+$-ions.}
	\label{fig:spectra}
\end{figure}
\section{\label{sec:Results}Results and discussion}
\subsection{Nanosecond spectroscopy}
In the present article we study He nanodroplets doped with Rb$_2$ molecules excited into electronic states at the laser wavelengths in the range 465-495~nm and 412-427~nm (20200-21500 and 23400-24300~\cm). At these wavelengths, the Rb$_2$ molecules are photoionized by resonant 1+1 one-color two-photon ionization. The measured yield of Rb$_2^+$ molecular ions created by the nanosecond (ns) laser is shown in Fig.~\ref{fig:spectra} b). Two maxima are clearly visible at 20500~\cm~and at 21170~\cm, and a third less pronounced peak appears at 23750~\cm. 

\begin{figure}
	\centering
	\includegraphics[width=0.8\textwidth]{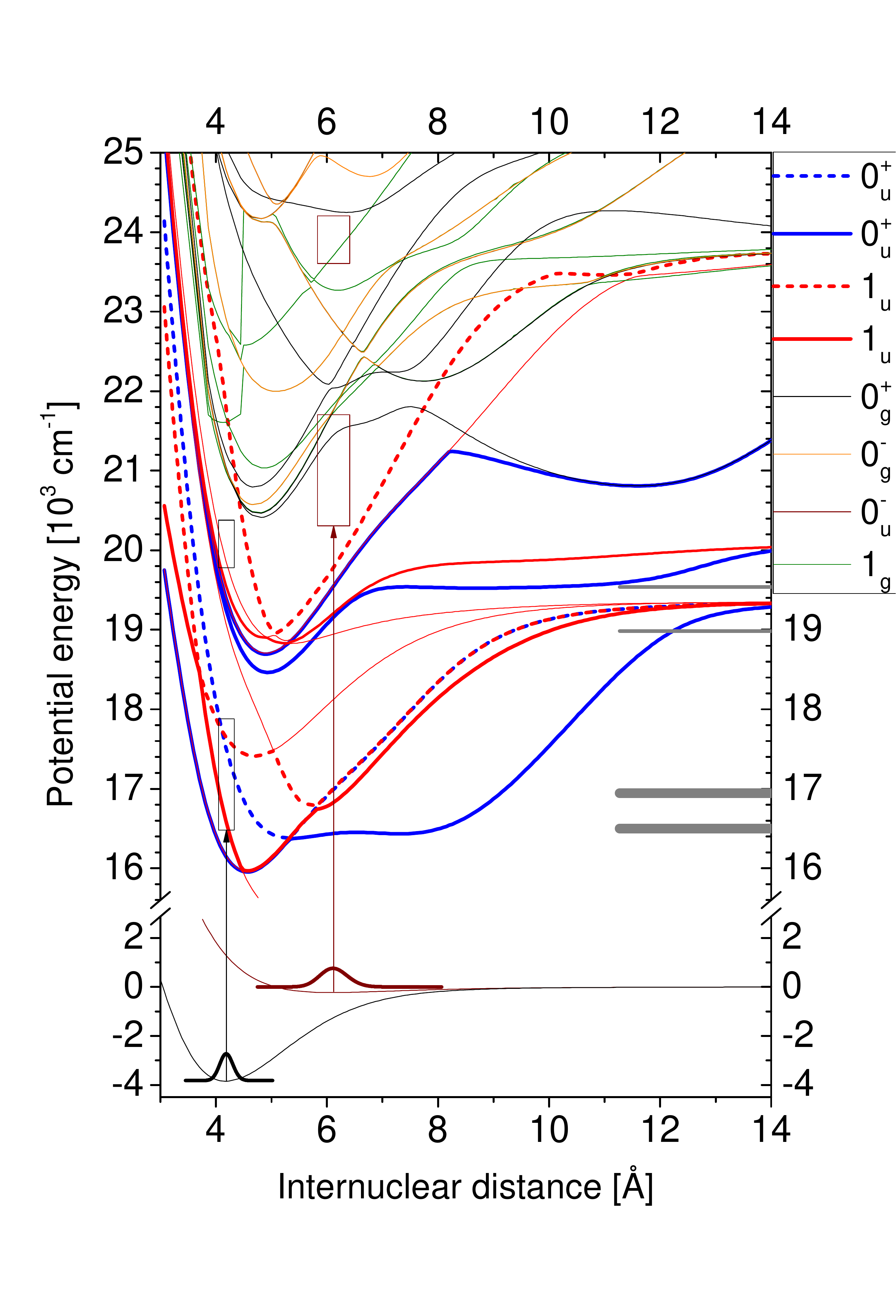}
	\caption{Potential energy curves of spin-orbit coupled Rb$_2$ states~\cite{Lozeille:2006}. The vertical arrows and rectangles indicate the range of laser excitation from the $^1\Sigma_g^+\left(0^+_g\right)$ and the $^3\Sigma_u^+\left(0^-_u\right)$ groundstates into the excited state manifold. For the excitation from $0^+_g\left(0^-_u\right)$ all relevant $u$ ($g$)-states are shown. The thick lines indicate the relevant curves for the present experiments. To distinguish states of the same $\Omega_{g,u}$-symmetry at avoided crossings some curves are drawn as dashed and some as solid. The thick gray horizontal lines on the right hand side show the calculated asymptotic energies inferred from the experimental KER spectra.}
	\label{fig:potentials}
\end{figure}
To assign these spectral features to molecular excitations we first inspect the Rb$_2$ spin-orbit coupled potential energy curves~\cite{Lozeille:2006} neglecting the influence of the droplet environment. The potential curves of $\Omega=0^{+/-}_{u,g}$ and $1_{u,g}$ symmetry accessible by optical transition from the lowest singlet or triplet states in the relevant wavelength range are depicted in Fig.~\ref{fig:potentials}. For the excitation from the $^1\Sigma_g^+\left(0^+_g\right)$ groundstate only ungerade states and from the lowest triplet state $^3\Sigma_u^+\left(0^-_u ,\, 1_u\right)$ only gerade states are accessible. States of $2_{u,g}$ symmetry cannot be excited out of the 0$^{+/-}_{g,u}$ groundstates~\cite{Lee:2000}. The vertical arrows and rectangles indicate the energy range covered by our experiments. In this intermediate range of excitation the energy levels are quite dense. This holds even when taking into account that vibronic lines of Ak molecules attached to He droplets are typically blue-shifted by up to about 100~\cm ~with respect to the gasphase values due to the interaction with the droplet environment~\cite{Stienkemeier2:1995,Mudrich:2004,Allard:2006,Auboeck:2010,Buenermann:2007,LoginovPRL:2011,LoginovPhD:2008}.


When spin-orbit-coupling is negligible (Hund's coupling case a), only singlet (triplet) states can be reached by (optical) excitation from a singlet (triplet) state. In the presence of strong spin-orbit coupling (Hund's case c) which is the case for heavy Ak dimers like Rb$_2$~\cite{Lozeille:2006}, this selection rule is relaxed. We note that the mixing of singlet and triplet states of Rb$_2$ due to spin-orbit coupling has previously been observed in several spectroscopic studies~\cite{Auboeck:2007,Auboeck:2010,Ernst:2006,Lozeille:2006,Park:2001}.


We therefore calculate the transition probabilities (Franck-Condon factors, FCF) for the allowed transitions from the lowest (Hund's case c) electronic levels $^1\Sigma_g^+ (0_g^+)$ and $^3\Sigma_u^+ (0^-_u)$ into excited spin-orbit states $\Omega=0_u,\, 1_u$ using R. LeRoy's program LEVEL~\cite{level}.
In the following, we neglect excitations out of the lowest triplet state $^3\Sigma_u^+$ because of the lacking correspondence of the respective FCF contours with the measured spectrum.
The calculated FCF for transitions out of the $^1\Sigma_g^+$ groundstate are shown in Fig.~\ref{fig:spectra} a) to compare with the measured photoionization spectrum in b). Each of the $\Omega=0_u,\, 1_u$-states is composed of different non-spin-orbit states in various ranges of the interatomic distance, as indicated. Crossings of coupled potential curves become avoided when taking spin-orbit-coupling into account. However, in wide regions of internuclear distance and energy the coupled potential curves still retain a dominant character of non-spin-orbit states. The dominant $\Sigma$ and $\Pi$-characters in the region of excitation are highlighted by the red and blue colors of symbols in Fig.~\ref{fig:spectra} a), respectively. 

The first broad peak around 20500~\cm~matches the absorption profile of the $1_u$-state. Without spin-orbit coupling this state would have  $^3\Sigma_u^+$-symmetry and dissociate towards the $5s+5p$ atomic asymptote. Due to the SO-coupling with a $^1\Pi_u$ and a $^3\Pi_u$ state, the latter state becomes binding with $1_u$ (or $0^+_u$)-symmetry asymptotically correlating to $5s+4d$.


The maximum around 21170~\cm~has overlap with the computed absorption profiles of a binding $^1\Sigma_u^+$($0^+_u$) and a $^1\Pi_u$($1_u$) state, both correlating to $5s+4d$. However, the measured peak is slightly red-shifted by 170 and 300~\cm~, respectively, compared to the computed profiles. 
Note that absorption features of intermediate excited states of other Ak atoms attached to He droplets have also been found to be shifted in the same range of wavenumbers with respect to simple model calculations~\cite{Buenermann:2007,Loginov:2011,Hernando:2012,Fechner:2012}. Most likely reasons for this deviation which are hard to quantify are inhomogeneous broadenings due to droplet size-dependent line shifts in combination with a broad droplet size distribution as well as due to local fluctuations of the He density around the dopant~\cite{Loginov:2011}. Besides, the accuracy of potential curves of such intermediate levels of excitation may be an issue. The left wing of the feature around 23800~\cm~overlaps with the right wing of the FCF contour of the $\Pi_u(1_u)$-state. At higher photon energies the feature reaches out to the FCF contours of the $^3\Sigma_u^+ (1_u,\,0_u^-)$-states. 

Another possible source of distortions of the spectra is the fragmentation of Rb$_3$ and larger clusters into Rb$_2$ which might add signal to the Rb$_2$ spectrum. However, given the low Rb vapor pressure in the doping cell as well as the lack of any signals of unfragmented Rb$_3^+$ up to the noise level of 0.03\% of the Rb$_2^+$-signal in the entire tuning range of our lasers, we dismiss fragmentation of heavier masses as highly unlikely for the conditions set in our experiment. While the He environment is likely to induce blue-shifting and additional broadening of the measured spectral features of the order of tens to a few hundreds of~\cm, we refrain from attempting to include such effects due to the poor characterization of the droplet effect on Ak metal molecules excited to intermediate states.

\begin{figure}
\centering
\includegraphics[width=0.6\textwidth]{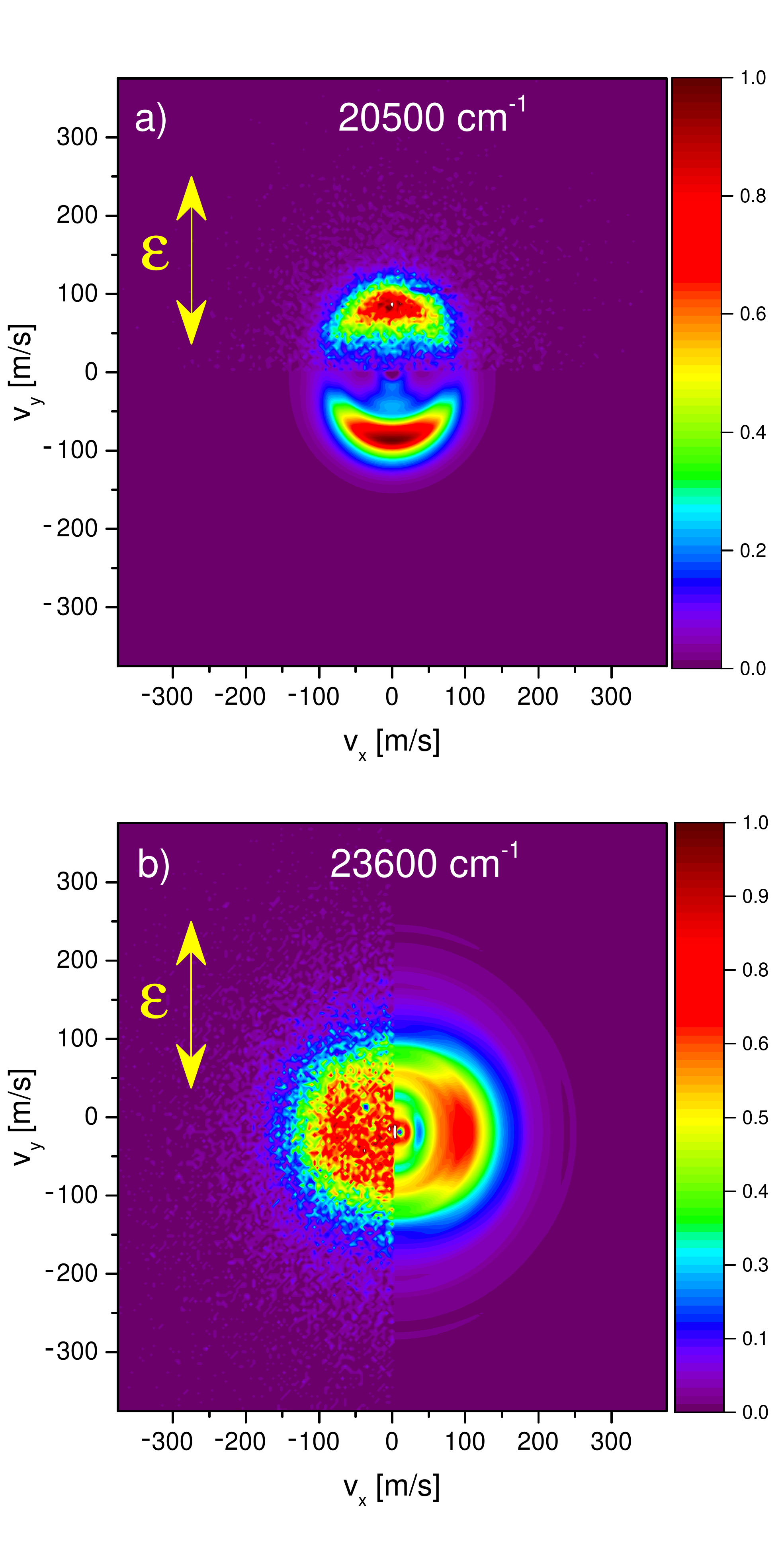}
\caption{Typical raw and inverted Rb$_2^+$ ion images recorded at two different wavenumbers of the dye laser. The vertical arrow indicates the laser polarization.}
\label{fig:images}
\end{figure}
\subsection{Rb$_2^+$ ion imaging}
More detailed information about the nature of the excited Rb$_2$-states is obtained from the velocity distributions of the desorbed Rb$_2^+$ ions measured by means of VMI using the dye laser. Typical VMIs recorded at the laser wavelength 488~nm (20500~\cm) and 424~nm (23600~\cm) are displayed in Fig.~\ref{fig:images} a) and b), respectively. The figures are made up of one half of the raw image [upper in a), left in b)] and one half of the Abel inverted image [bottom in a), right in b)]. Clearly, the images display angular distributions with pronounced opposite anisotropies, where in a) the preferred direction of Rb$_2$ emission is along the polarization of the laser (yellow arrow), whereas in b) it is perpendicular to the laser polarization. 

\begin{figure}
	\centering
	\includegraphics[width=0.6\textwidth]{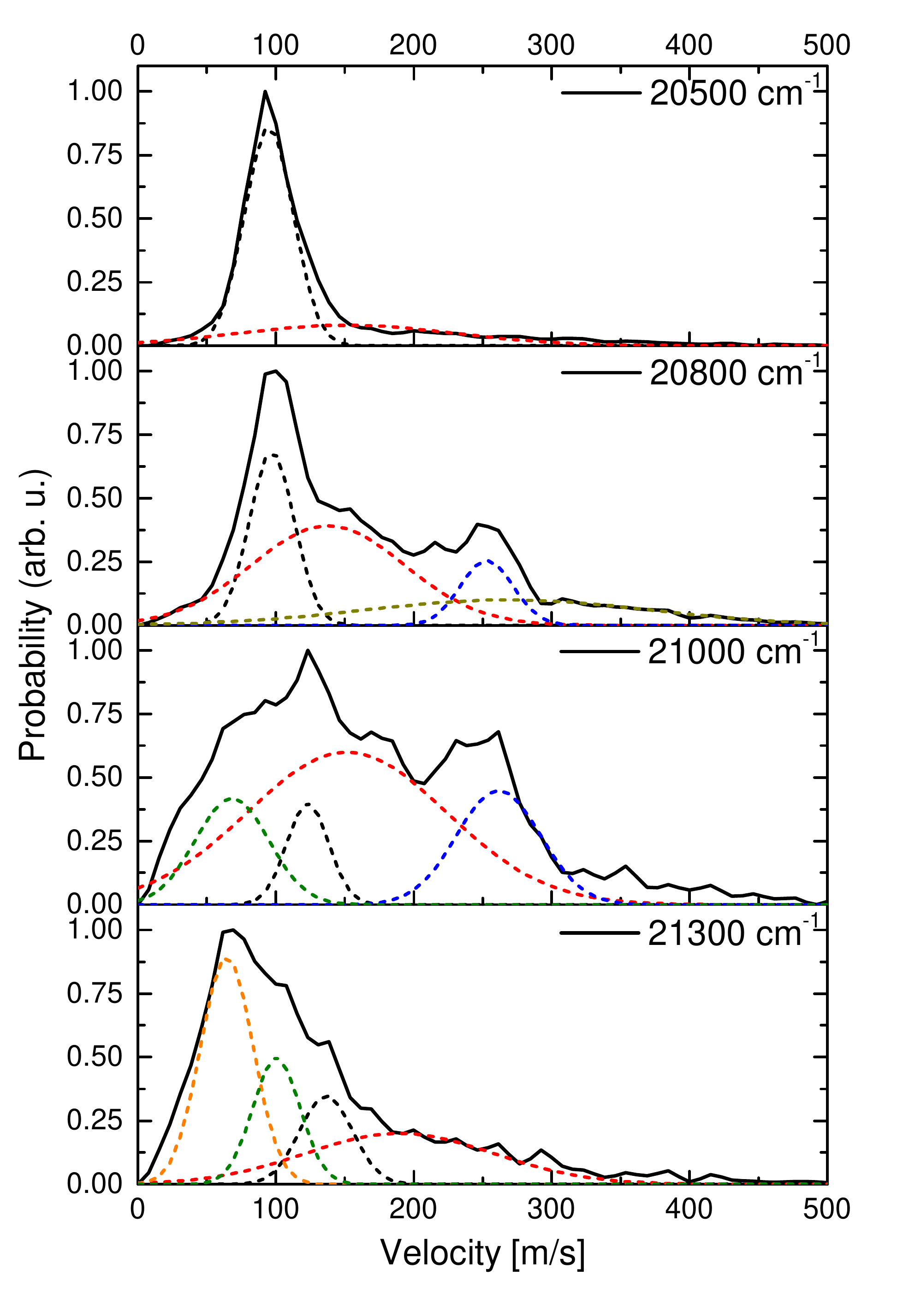}
	\caption{Rb$_2^+$ velocity distributions inferred from VMIs recorded at various wavenumbers of the dye laser. The dashed colored lines indicate the results of a fitting procedure using the sum of 2 to 4 Gaussian functions.}
	\label{fig:speed}
\end{figure}
Aside from angular distributions, from these VMIs we can infer the final velocity of desorbed Rb$_2$ by 
integrating the Abel inverted images over angles. The resulting velocity distributions for various laser wavelengths are shown in Fig.~\ref{fig:speed}. In contrast to the velocity distributions of desorbed Rb$^+$ atomic ions measured at similar laser wavelengths~\cite{Fechner:2012,Vangerow:2014}, the Rb$_2^+$ velocity distributions feature more complex structures. They are best modeled by a sum of two to four Gaussian functions, depicted as smooth colored lines in Fig.~\ref{fig:speed}. The presence of several velocity components indicates that either a superposition of initial states is excited, where each state evolves along a distinct potential energy curve with respect to the interaction of the excited Rb$_2$ with the He droplet, or that relaxation occurs due to the coupling of the initial to some final states, probably induced by the He droplet environment. Such behavior was recently observed for Na atoms excited into high-lying levels which were subjected to massive droplet-induced configuration interaction~\cite{Loginov:2015}. 


The angular anisotropy of velocity distributions resulting from the dissociation of a diatomic molecule by one-photon absorption is quantified by the parameter $\beta$~\cite{Zare:1972,Hernando:2012,Fechner:2012}. Here, $\beta=2$ specifies a transition where the dipole moment points parallel to the molecular axis. This applies to \textit{parallel} transitions with no change of angular momentum projection, such as $\Sigma\rightarrow\Sigma$. The resulting angular distribution of fragments is peaked along the polarization direction, similar to that of Fig.~\ref{fig:images} a). In contrast, for \textit{perpendicular} transitions of the type $\Sigma\rightarrow\Pi$ fragmentation occurs preferentially in the plane perpendicular to the polarization as the transition dipole moment points perpendicular to the molecular axis and therefore $\beta=-1$. In that case, VMIs resemble the one shown in Fig.~\ref{fig:images} b).

\begin{figure}
\centering
\includegraphics[width=0.6\textwidth]{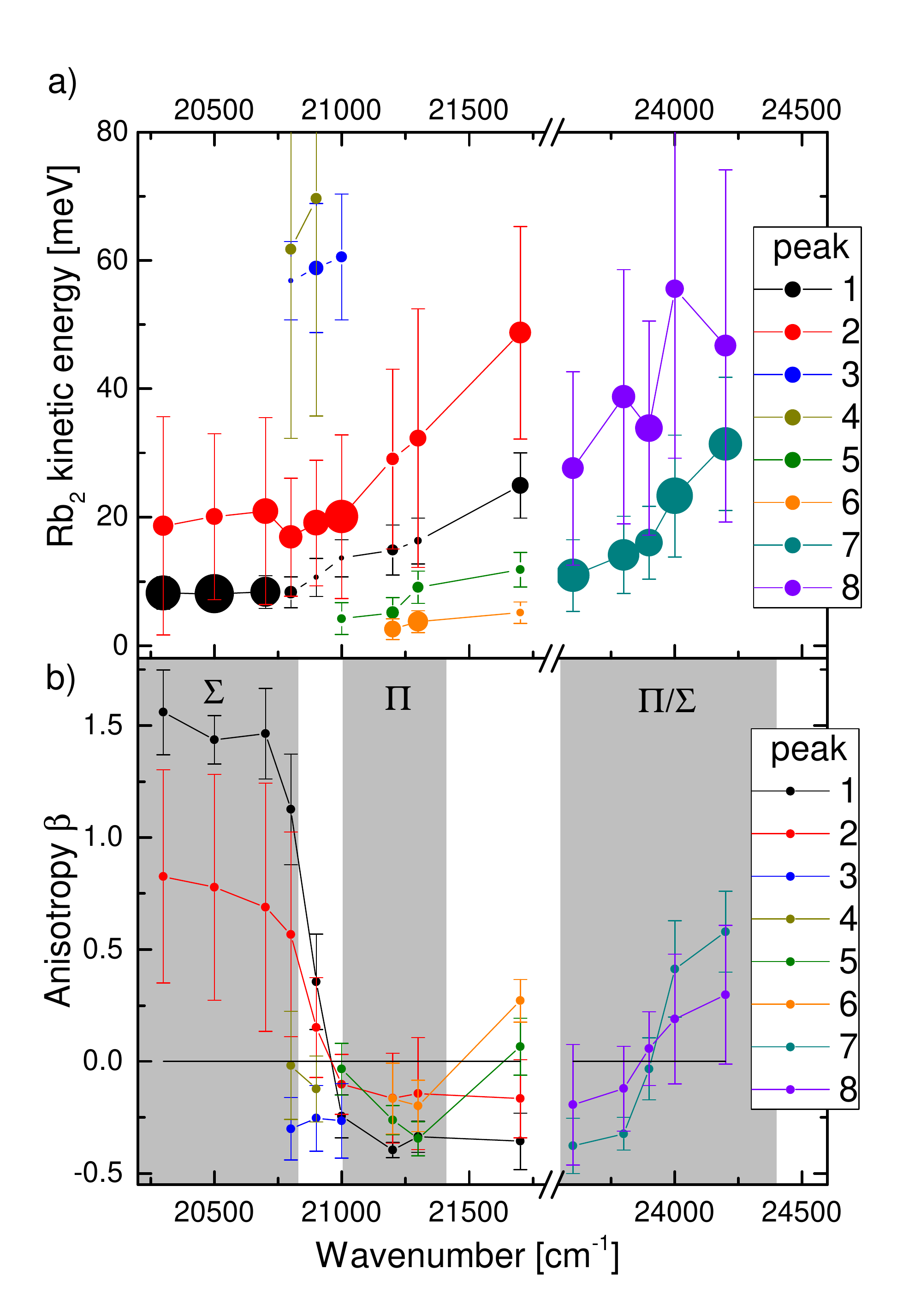}
\caption{a) Graphical compilation of Rb$_2^+$ kinetic energies resulting from fits of the velocity distributions. The sizes of the circles indicate the relative weights of the respective velocity components. b) Anisotropy parameter $\beta$ associated with the individual velocity components.}
\label{fig:KERbeta}
\end{figure}
The mean values of Rb$_2^+$ kinetic energies and anisotropies obtained from the fits of the velocity distributions are graphically summarized in Fig.~\ref{fig:KERbeta} for various investigated laser wavelengths. The symbol colors match the fit curves of the velocity distributions shown in Fig.~\ref{fig:speed}. The error bars reflect the velocity ranges used to calculate the weighted mean values and approximately match the full widths at half maxima (FWHM) of the gaussian fit curves. Thus, overlapping error bars indicate that the corresponding peaks partly or even fully overlap. The sizes of the circles visualize the relative peak integrals of the fit curves. 

The velocity distributions in the range 20300~cm$^{-1}$ to 20700~cm$^{-1}$ are nearly constant featuring one large (black) and one small peak (red). By increasing the photon energy to 20800~cm$^{-1}$ additional components at higher velocities appear (blue and dark yellow). New components at lower velocities (green and orange) grow in at 21000~cm$^{-1}$, while others (dark yellow and blue) disappear. These four peaks subsequently shift towards higher velocities as the laser is tuned to 21700~cm$^{-1}$. At higher wavenumbers 23500-24400~cm$^{-1}$ the distributions have two features which shift to higher velocities with increasing wavenumber.

For every gaussian fit curve we calculate the weighted average of the $\beta$-parameter, which is depicted with the same color-coding in Fig.~\ref{fig:KERbeta} b). The error bars illustrate the variation of $\beta$ within the FWHM. Since in the studied regions of excitation the excited states are predominantly determined in their character by one single non-SO-coupled state we discuss the recorded angular distributions in terms of these uncoupled states. 

The lower part of the tuning range (20300-20800~cm$^{-1}$) shows an anisotropy parameter $\beta> 1$ ($\beta =1.6(2)$-$1.1(2)$ for peak 1 and $\beta=0.8(5)$-$0.6(5)$ for peak 2) which means that predominantly a parallel transition is driven. The accessible $1_u$-state correlating to the $5s+5p$-atomic asymptote has predominantly $^3\Sigma_u^+$-symmetry in the region of excitation. Thus, a $\Sigma-\Sigma$ transition is excited which leads to the observed anisotropy of the velocity distribution in the direction of the laser polarization. The fact that the Rb$_2$ kinetic energy is constant in this range of photon energies is at odds with previous measurements using Ak metal atoms where linearly increasing kinetic energies with increasing photon energy were found~\cite{Hernando:2012,Fechner:2012,Vangerow:2014}. Those experiments were interpreted in terms of the pseudo-diatomic model in which the repulsive dopant-He droplet interaction energy is converted into kinetic energy released in the course of the desorption process. Accordingly, the observed constant and rather small kinetic energy of desorbed Rb$_2$ points at weak Rb$_2$-He droplet repulsion in that excited state and at inefficient conversion of internal vibrational energy of the Rb$_2$ into kinetic energy.

In the range from 20900 to 21700~\cm~peak 2 gives the largest contribution. The mean anisotropy of $\beta=-0.13(2)$ points at an excited $\Pi$-state contribution. This agrees with the FCF calculation according to which the $2^1\Pi_u$-state carries a larger transition strength than the $2^1\Sigma_u^+$-state. The linear dependence of Rb$_2^+$ kinetic energy as a function of wavenumber in this range of photon energies indicates that now potential energy, either stored in internal Rb$_2$ excitations or in the Rb$_2$-He$_N$ inter-cluster degree of freedom, induces repulsion of Rb$_2$ with respect to the droplet surface and therefore partly converts to kinetic energy of the desorbing Rb$_2$. The extrapolation of a linear fit function to the data down to zero kinetic energy then yields the internal energy of the free Rb$_2$ molecule in that excited state~\cite{Hernando:2012,Fechner:2012,Vangerow:2015}. The found value of about 20360~\cm, which is displayed as the bottom thick gray line on the right hand side of Fig.~\ref{fig:potentials}, roughly matches the bottom of the potential well of the lowest $0_u^+$($2^1\Sigma_u^+$)-state. Thus, we assume that in the course of desorption both the Rb$_2$-He droplet interaction energy as well as Rb$_2$ vibrational energy are converted into kinetic energy.

From the slope of the linear fit we deduce an effective mass of the He droplet $m_{eff}=66.2$~a.\,u., i.\,e. a number of $16.6$ He atoms interacts with the excited Rb$_2$ in the desorption process. This is about twice the value previously found for Rb atoms ($\sim 10$)~\cite{Vangerow:2014} which appears plausible as the Rb$_2$ dimer has roughly twice the size of the Rb atom.
Linear regressions of peaks 5 and 6 extrapolate to a potential energy at vanishing Rb$_2$ kinetic energy in the range 20710-20825~\cm, which is marked by the upper thick gray line on the right hand side of Fig.~\ref{fig:potentials}. This energy matches the potential well at the internuclear distance $R\approx 6$~\AA~of the upper $1_u$-state correlating to the $5s+5p$ atomic asymptote (dashed red line). In the range of $R$ where this state is excited it is dominated by the $^1\Pi_u$-contribution which is consistent with the measured negative anisotropy of $\beta=-0.23(3)$ (peak 5) and $\beta=-0.26(9)$ (peak 6). 


In the laser wavenumber range 23600-24200~\cm~the Rb$_2$ kinetic energy distributions are best fitted by a sum of two gaussian functions, peaks 7 and 8 in Fig.~\ref{fig:KERbeta}. The anisotropy of both components evolves from $\beta=-0.38$ (peak 7) and  $\beta=-0.19$ (peak 8) to $\beta=0.58$ and $\beta=0.30$, respectively, for increasing wavenumber. This points at this spectral region being located in the crossover region from $\Sigma-\Pi$ to $\Sigma-\Sigma$ transitions. This conclusion is supported by the FCF spectrum in Fig.~\ref{fig:spectra} which features the $4^1\Pi_u$-state at the lower and the $3^3\Sigma_u^+$-state at the upper end of this range of wavenumbers. 

From linear regressions we extrapolate to the minimum potential energy of the state associated with peak 8 of about 22800~\cm~which matches the potential well depth of the upper $4^1\Pi_u (1_u)$-state  (see Fig.~\ref{fig:potentials}).
Peak 7 extrapolates to a minimum energy of about 23360~\cm~which can be assigned to the outer plateau of the $3^3\Sigma_u^+$-state potential curve correlating to the $5s+6p$-atomic asymptote. Thus, again it appears that the Rb$_2$ internal energy is efficiently converted into kinetic energy release by droplet-induced relaxation, where a fraction of about 58\% of molecules stabilizes in the plateau of the $3^3\Sigma_u^+$-potential and the remainder relaxes down into the wells of the $4^1\Pi_u$ or $3^3\Sigma_u^+$-potentials. The slopes of the linear fits yield effective numbers of He atoms interacting with the desorbing excited Rb$_2$ molecules of 17 and 18 for peaks 7 and 8, respectively. 

\begin{figure}
\centering
\includegraphics[width=0.7\textwidth]{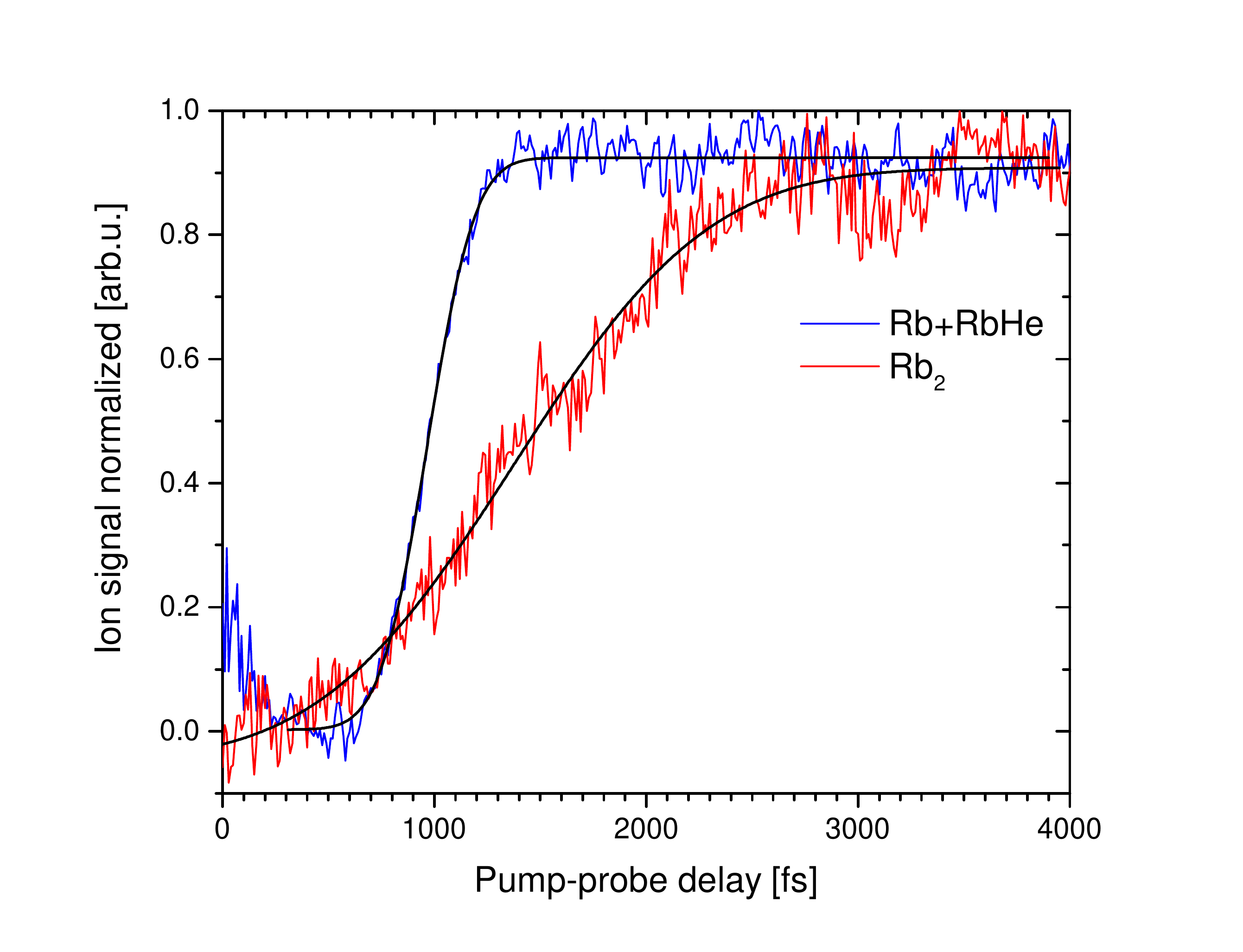}
\caption{Femtosecond pump-probe measurements of the yields of Rb$^+$ and RbHe$^+$ ions in a), and Rb$_2^+$ in b), recorded at a laser wavelength 413~nm (24210~\cm ). The smooth black line is a fit of an error function.}
\label{fig:ppions}
\end{figure}
\subsection{\label{sec:Femtosecond}Femtosecond desorption dynamics}
In addition to measuring the yields and final velocity distributions of Rb$_2^+$ ions generated by ns laser ionization, we have investigated the time-evolution of the Rb$_2^+$ signals in the course of the desorption process using fs pump-probe spectroscopy. Fig.~\ref{fig:ppions} shows the transient yield of Rb$_2^+$ ions recorded at a laser wavelength of 413~nm (242100~\cm ), which we compare to the sum of the yields of Rb$^+$ and RbHe$^+$ ions recorded at the same wavelength. In these measurements, a first pump pulse resonantly excites the dopant atom or molecule while ionization is induced by a second time-delayed probe pulse.

At this wavelength, the Rb atomic dopant is excited into the $6p$-state which is perturbed by the He droplet to form pseudo-diatomic $6p\Sigma$ and $6p\Pi$ states~\cite{Vangerow:2015}. Rb$_2$ dimers are excited in between the $^1\Pi_u$ (1$_u$) and $^3\Sigma_u^+$ (1$_u$,0$_u^-$) states correlating to the $5s+6p$ atomic asymptote. The observed time-delayed rise of the ion signals reflects the competing dynamics of desorption of the excited dopant off the He droplet surface and the solvation of the ionized dopant into the He droplet interior induced by attractive He-ion interactions. Thus, at short pump-probe delay ionization of the dopant occurs in close vicinity of the He droplet such that all ions turn over and fall-back into the He droplet to form stably bound snowball complexes~\cite{Theisen:2010,Vangerow:2015}. Only at larger delays the excited dopants have gained enough kinetic energy so as to overcome the attraction they experience after ionization by the probe pulse. 

In Fig.~\ref{fig:ppions} we plot the sum of Rb$^+$ and RbHe$^+$ ions as the characteristic signal for the desorption dynamics of the atomic dopants. Upon excitation of a Rb atom on the He droplet surface stable RbHe excited molecules, so-called exciplexes, can form and decay in the course of desorption off the droplet~\cite{Vangerow:2015}. This may lead to transient redistribution of Rb$^+$ and RbHe$^+$ signals irrelevant for the present study.
By fitting the transient yield curves with an error function we obtain the fall-back times $t_c^{Rb}=0.968(3)$~ps and $t_c^{Rb2}=1.38(2)$~ps as those delay times at which the signal has increased to half the final value reached at long delays. The significantly longer fall-back time for Rb$_2^+$ points at weaker repulsion as compared to the excited Rb atom, in line with our expectation based on the trend that Ak clusters are more strongly bound to He droplets than atoms and that Pauli repulsion due to electronic excitation tends to be weaker for large molecules (see Sec.~\ref{sec:Intro}).
Thus, we predict that larger Rb oligomers will feature even longer fall-back times than Rb$_2$ and may not desorb at all for dopant clusters exceeding a certain size. Besides this, molecules and larger oligomers have larger masses and are likely to induce larger effective masses of the He droplet they interact with due to their larger spatial extension, which additionally slows down the desorption process. 

\begin{figure}
\centering
\includegraphics[width=0.7\textwidth]{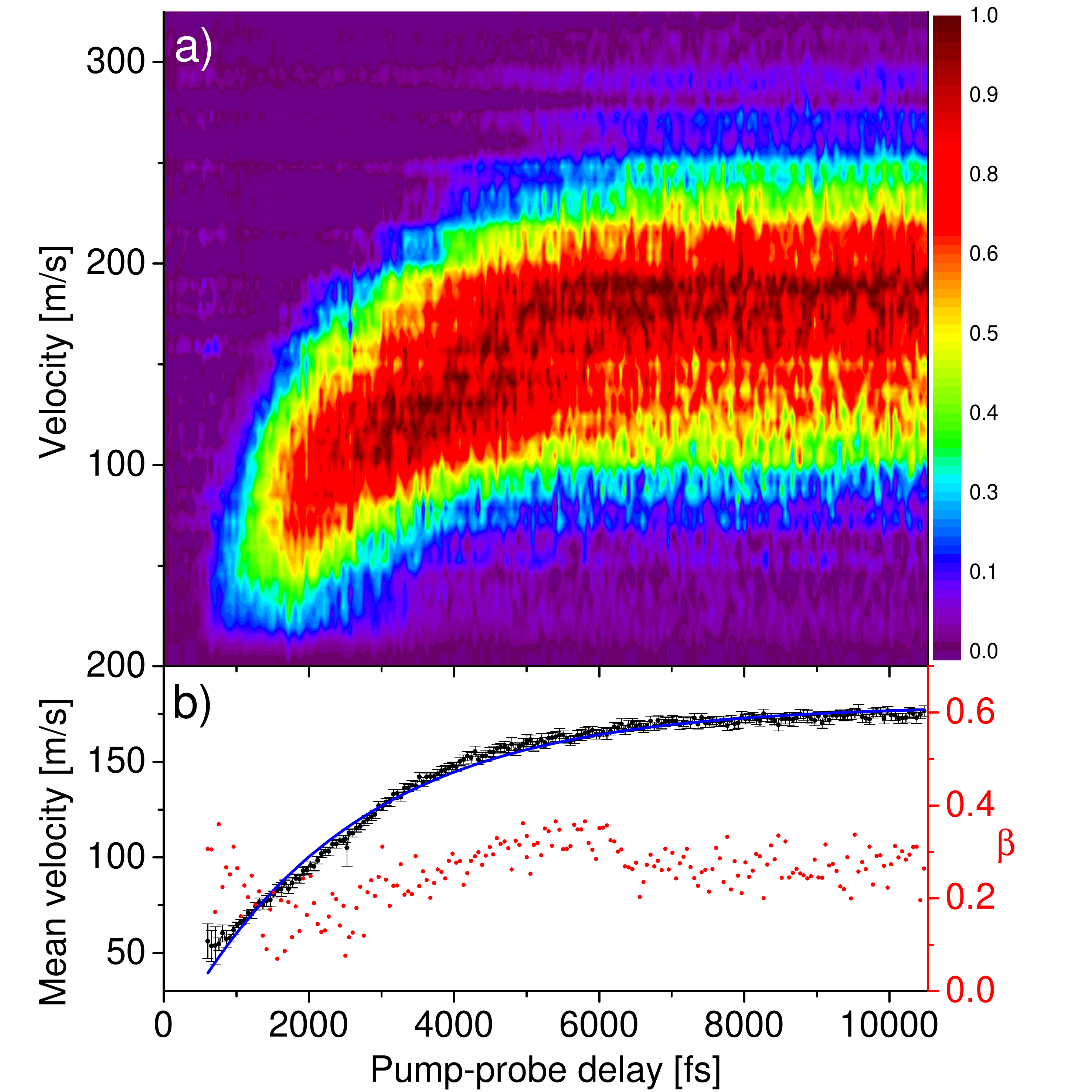}
\caption{a) Time evolution of Rb$_2^+$ ion speed distributions inferred from velocity-map images recorded as a function of pump-probe delay. Each distribution is background subtracted and weighted with the corresponding ion yield (see text). b) Transient mean velocities (black symbols, left vertical scale) and anisotropy parameters $\beta$ (red symbols, right vertical scale). The blue line is a fit of an exponential function to the velocity data.}
\label{fig:ppspeed}
\end{figure}

Apart from transient ion yields, velocity map images of Rb$_2^+$ fragments have been recorded for different pump-probe delays. Fig.~\ref{fig:ppspeed} a) shows the speed distributions obtained by inverse Abel transformation of the raw images. For each pump-probe delay step, the maximum of the spectrum is normalized to the corresponding ion yield. In addition to the pump-probe correlated signal, a delay independent background contribution of about ~15\% of the signal at long delays is measured. In order to extract the pump-probe correlated data, this contribution is subtracted. The measured velocity distributions are rather broad and show a slight asymmetry. Nevertheless, a shift of the peak position towards higher velocities with increasing delay is clearly visible. To analyze this in more detail, each spectrum is fitted by a epsilon skew gaussian distribution~\cite{Mudholkar:1999}, which was found to yield the best fit result. From the fit result at every delay step the mean velocity and the anisotropy parameter are inferred and plotted in Fig.~\ref{fig:ppspeed} b). The anisotropy parameter is calculated by weighting the angular distribution with the velocity probability distribution within one standard deviation. 
 
	
When significant Rb$_2^+$ ion signal starts to appear in the ion images at a pump-probe delay of about 600~fs we measure velocities with a mean value of 54(3)~m/s. This time marks the onset of the pump probe-correlated signal within the experimental uncertainty. It is identified with the time when a first fraction of the created wavepacket has gained enough kinetic energy to overcome the potential barrier provided by the ionic potential curve. The origin of the leveling out of the ion velocity at short delay to a finite value around 55~m/s is unresolved at this point. Note that experimental artifacts such as a limited resolution of the imaging setup can be excluded. We mention that similar finite velocity values at short delay are also observed in for Rb$^+$ and RbHe$^+$ ion in experiments where the He droplets are doped with single Rb atoms. This will be discussed in a forthcoming paper.

With increasing delay the velocity shifts to higher values till it reaches 90$\%$ of the asymptotic value of 179.6(5)~m/s at a delay of $5.65(8)$~ps. This monotonous increase in velocity directly reflects the dynamics of desorption of excited Rb$_2$ molecules off the He surface. As expected, at long delay times the velocity map ion images converge towards those measured using the ns laser within the experimental uncertainty.  

From fitting an exponential function $v(t)=v_{rise}\left( 1-\exp\left(-\frac{t}{\tau}\right)\right)$ to the data we obtain a characteristic time constant for the desorption process $\tau=2.45(2)$~ps. Note that this value exceeds the Rb$_2$ fall-back time derived from the ion yield measurement. The difference between ion yield and velocity transients is attributed to the more extended range of repulsive interaction of the excited Rb$_2$ molecule as compared to the range of attraction of the Rb$_2^+$ ion towards the He droplet. As a consequence, the excited Rb$_2$ continues to be accelerated away from the droplet surface even beyond the fall-back time.  

The time evolution of the anisotropy parameter $\beta$ is shown as red symbols in Fig.~\ref{fig:ppspeed}. It only slightly varies between $\beta=0.18(2)$ at short delays and $\beta=0.28(1)$ at long delays. The latter value falls slightly below the value measured with the ns laser ($\beta =0.6(2)$). This is likely due to the excitation of a superposition of states of opposing symmetry by the broad-band fs laser where the $\Sigma$-component dominates.

\begin{figure}
	\centering
	\includegraphics[width=0.6\textwidth]{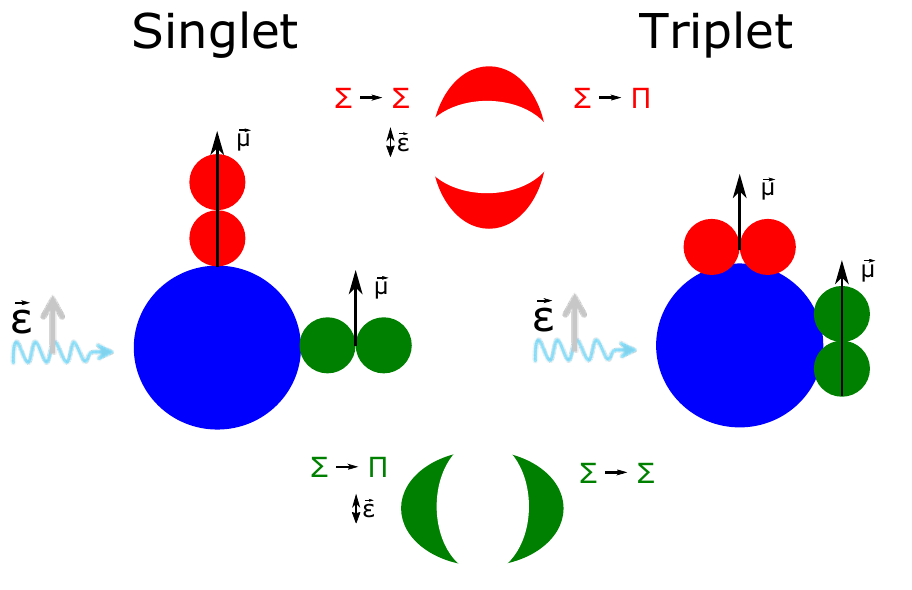}
	\caption{Schematic representation of the configuration of Rb$_2$ molecules in the singlet (left) and triplet (right) groundstates attached to He droplets and the corresponding angular distributions upon desorption. Vector $\vec{\varepsilon}$ indicates the direction of laser polarization and $\vec{\mu}$ is the transition dipole moment. See text for details.}
	\label{fig:diagram}
\end{figure}
\subsection{\label{sec:Discussion} Desorption model}
First of all we point out that our time- and velocity-resolved measurements confirm our conception of prompt, pseudo-diatomic dissociation of the Rb$_2$He$_N$ complex as discussed above in the context of our ns experiments. The observed pump-probe delay-dependent dynamics clearly shows that in the studied energy range one-photon vibronic excitation of neutral Rb$_2$ dimers induces an accelerated, directed motion of the Rb$_2$ away from the droplet surface during a few ps. 

Based on these experimental findings we suggest the following general scheme for the desorption process of Rb$_2$ diatomic molecules from the surface of He droplets. The argumentation most likely holds for other Ak metal dimers as well. The central assumption is that the quantization axis which determines the symmetry of the molecular state and thus the direction of the transition dipole moment is determined by the molecular axis of the Rb$_2$ dopant. This is justified by the fact that the excitation is mostly centered on the Rb$_2$ due to droplet excitations falling into very different ranges of energy as compared to those of the Rb$_2$ molecules. Furthermore, the excited states are governed by the generally much stronger intramolecular interactions ($\sim 100$-$1000$~\cm) as compared to the interaction of the Rb$_2$ dimer with the He droplet ($\sim 10$-$100$~\cm). 

Thus, for the singlet groundstate where the Rb$_2$ stands perpendicular on the He droplet surface~\cite{Bovino:2009,Leino:2011}, the angular distribution of Rb$_2$ ``fragments'' coincides with that of Rb atomic fragments for the hypothetical case that free Rb$_2$ molecules were excited into dissociative states of the same symmetry. This situation is schematically sketched on the left hand side of Fig.~\ref{fig:diagram}. In contrast, the lowest triplet state of Rb$_2$ would result in an angular distribution of desorbed Rb$_2$ molecules of the opposite anisotropy as compared to the singlet groundstate. That is, counter-intuitively, $\beta=2$ would correspond to the \textit{perpendicular} transition $\Sigma\rightarrow\Pi$, as shown on the right hand side of Fig.~\ref{fig:diagram}. Unfortunately this conjecture cannot be directly assessed with the data measured in this work. However, in future experiments we will check this situation by exciting better isolated, low-lying triplet states of Ak dimers as we have done before~\cite{Claas:2007,Mudrich:2009}. In addition to the angular distributions, the internal states of the Rb$_2$ molecule appear to govern the kinetic energy released into the translational degree of freedom of the desorbing Rb$_2$ off the droplet surface through efficient coupling between intramolecular vibration and Rb$_2$-He$_N$ relative motion. The observed desorption timescale of a few picoseconds being in the same range as the vibrational periods of the Rb$_2$ dimer states is in agreement with this conclusion. In contrast, previous experiments indicated that Rb$_2$ dimers remain attached to the droplet after excitation to the lowest lying triplet state even for high vibrational states\cite{Mudrich:2009,Gruner:2011}. Thus, we argue that the desorption probability most strongly depends on the electronic state. The analysis of time-resolved imaging experiments with Rb atoms attached to He droplets excited to different electronic states will further clarify this issue.  

\section{\label{sec:Summary}Summary and conclusions}
The present study clearly demonstrates that Rb$_2$ dimers formed on the surface of He nanodroplets promptly desorb off the surface upon electronic excitation into intermediate states, similarly to Rb and other alkali metal dopant atoms. However, in contrast to Rb atoms, the angular distribution of detected molecular ions is not determined by the symmetry of the dopant-droplet complex, but rather by the symmetry of the internal molecular states of the Rb$_2$ dopant. The latter remains weakly perturbed due to the relatively weak Ak-He coupling. We conclude that this leads to opposite anisotropies of the angular distributions of desorbed Rb$_2$ formed in the singlet and triplet groundstates when driving transitions of the same symmetry. 

Likewise, the kinetic energy distributions of desorbed Rb$_2$ appear to be mostly determined by the internal energy of Rb$_2$ which is converted into kinetic energy released into the translational motion of the desorbing Rb$_2$. This is enabled by efficient He droplet-induced vibrational relaxation and coupling of intra and inter-molecular degrees of freedom of the Rb$_2$He$_N$ complex. Therefore the resulting multi-peaked structures of kinetic energy distributions elude from a simple interpretation in terms of pseudo-diatomic dissociation as in the case of Ak metal atoms on He droplets. 

Femtosecond time-resolved measurements of Rb$^+_2$ ion yields and velocities reveal qualitatively the same transient dynamics as Rb atoms, which is determined by the competition of repulsion of the excited dopant away from the droplet surface and attraction of the dopant towards the droplet once ionized. However, this desorption dynamics proceeds more slowly for Rb$_2$ dimers as for Rb atoms at the probed laser wavelength pointing at a weaker repulsion of excited Rb$_2$ than of excited Rb. In future experiments we will refine this study by probing Ak metal dimers in better defined singlet as well as triplet states. Moreover, time-resolved imaging spectroscopy of Ak metal trimers~\cite{Giese:2011} and larger clusters~\cite{Schulz:2004,Droppelmann:2009} will shed new light on both the dynamics of intra-cluster degrees of freedom as well as the complex cluster-He droplet couplings.

\begin{acknowledgement}

The authors gratefully acknowledge support by the Deutsche Forschungsgemeinschaft in the frame of project MU~2347/6-1 as well as IRTG~2079. J. v. V. is supported by the Landesgraduiertenf{\"o}rderungsgesetz of Baden-W{\"u}rttemberg.

\end{acknowledgement}


\begin{mcitethebibliography}{68}
	\providecommand*\natexlab[1]{#1}
	\providecommand*\mciteSetBstSublistMode[1]{}
	\providecommand*\mciteSetBstMaxWidthForm[2]{}
	\providecommand*\mciteBstWouldAddEndPuncttrue	
	{\def\EndOfBibitem{\unskip.}}
	\providecommand*\mciteBstWouldAddEndPunctfalse
	{\let\EndOfBibitem\relax}
	\providecommand*\mciteSetBstMidEndSepPunct[3]{}
	\providecommand*\mciteSetBstSublistLabelBeginEnd[3]{}
	\providecommand*\EndOfBibitem{}
	\mciteSetBstSublistMode{f}
	\mciteSetBstMaxWidthForm{subitem}{(\alph{mcitesubitemcount})}
	\mciteSetBstSublistLabelBeginEnd
	{\mcitemaxwidthsubitemform\space}
	{\relax}
	{\relax}
	
	\bibitem[Krotscheck and Zillich(2001)Krotscheck, and Zillich]{Krotscheck:2001}
	Krotscheck,~E.; Zillich,~R. Dynamics of $^4\text{He}$ droplets. \emph{J. Chem.
		Phys.} \textbf{2001}, \emph{115}, 10161--10174\relax
	\mciteBstWouldAddEndPuncttrue
	\mciteSetBstMidEndSepPunct{\mcitedefaultmidpunct}
	{\mcitedefaultendpunct}{\mcitedefaultseppunct}\relax
	\EndOfBibitem
	\bibitem[Barranco \latin{et~al.}(2006)Barranco, Guardiola, Hern{\'a}ndez,
	Mayol, Navarro, and Pi]{Barranco:2006}
	Barranco,~M.; Guardiola,~R.; Hern{\'a}ndez,~S.; Mayol,~R.; Navarro,~J.; Pi,~M.
	Helium Nanodroplets: An Overview. \emph{J. Low Temp. Phys.} \textbf{2006},
	\emph{142}, 1--81\relax
	\mciteBstWouldAddEndPuncttrue
	\mciteSetBstMidEndSepPunct{\mcitedefaultmidpunct}
	{\mcitedefaultendpunct}{\mcitedefaultseppunct}\relax
	\EndOfBibitem
	\bibitem[Hernando \latin{et~al.}(2012)Hernando, Barranco, Pi, Loginov, Langlet,
	and Drabbels]{Hernando:2012}
	Hernando,~A.; Barranco,~M.; Pi,~M.; Loginov,~E.; Langlet,~M.; Drabbels,~M.
	Desorption of alkali atoms from $^4$He nanodroplets. \emph{Phys. Chem. Chem.
		Phys.} \textbf{2012}, \emph{14}, 3996--4010\relax
	\mciteBstWouldAddEndPuncttrue
	\mciteSetBstMidEndSepPunct{\mcitedefaultmidpunct}
	{\mcitedefaultendpunct}{\mcitedefaultseppunct}\relax
	\EndOfBibitem
	\bibitem[Mudrich and Stienkemeier(2014)Mudrich, and Stienkemeier]{Mudrich:2014}
	Mudrich,~M.; Stienkemeier,~F. Photoionisaton of pure and doped helium
	nanodroplets. \emph{Int. Rev. Phys. Chem.} \textbf{2014}, \emph{33},
	301–--339\relax
	\mciteBstWouldAddEndPuncttrue
	\mciteSetBstMidEndSepPunct{\mcitedefaultmidpunct}
	{\mcitedefaultendpunct}{\mcitedefaultseppunct}\relax
	\EndOfBibitem
	\bibitem[Ziemkiewicz \latin{et~al.}(2015)Ziemkiewicz, Neumark, and
	Gessner]{Ziemkiewicz:2015}
	Ziemkiewicz,~M.~P.; Neumark,~D.~M.; Gessner,~O. Ultrafast electron dynamics in
	helium nanodroplets. \emph{Int. Rev. Phys. Chem.} \textbf{2015}, \emph{34},
	239--267\relax
	\mciteBstWouldAddEndPuncttrue
	\mciteSetBstMidEndSepPunct{\mcitedefaultmidpunct}
	{\mcitedefaultendpunct}{\mcitedefaultseppunct}\relax
	\EndOfBibitem
	\bibitem[Stienkemeier and Lehmann(2006)Stienkemeier, and
	Lehmann]{Stienkemeier:2006}
	Stienkemeier,~F.; Lehmann,~K. Spectroscopy and dynamics in helium nanodroplets.
	\emph{J.~Phys.~B} \textbf{2006}, \emph{39}, R127 -- R166\relax
	\mciteBstWouldAddEndPuncttrue
	\mciteSetBstMidEndSepPunct{\mcitedefaultmidpunct}
	{\mcitedefaultendpunct}{\mcitedefaultseppunct}\relax
	\EndOfBibitem
	\bibitem[Loginov and Drabbels(2007)Loginov, and Drabbels]{Loginov:2007}
	Loginov,~E.; Drabbels,~M. Excited State Dynamics of Ag Atoms in Helium
	Nanodroplets. \emph{J. Phys. Chem. A} \textbf{2007}, \emph{111},
	7504--7515\relax
	\mciteBstWouldAddEndPuncttrue
	\mciteSetBstMidEndSepPunct{\mcitedefaultmidpunct}
	{\mcitedefaultendpunct}{\mcitedefaultseppunct}\relax
	\EndOfBibitem
	\bibitem[Smolarek \latin{et~al.}(2010)Smolarek, Brauer, Buma, and
	Drabbels]{Smolarek:2010}
	Smolarek,~S.; Brauer,~N.~B.; Buma,~W.~J.; Drabbels,~M. IR Spectroscopy of
	Molecular Ions by Nonthermal Ion Ejection from Helium Nanodroplets. \emph{J.
		Am. Chem. Soc.} \textbf{2010}, \emph{132}, 14086--14091\relax
	\mciteBstWouldAddEndPuncttrue
	\mciteSetBstMidEndSepPunct{\mcitedefaultmidpunct}
	{\mcitedefaultendpunct}{\mcitedefaultseppunct}\relax
	\EndOfBibitem
	\bibitem[Theisen \latin{et~al.}(2011)Theisen, Lackner, and Ernst]{Theisen:2011}
	Theisen,~M.; Lackner,~F.; Ernst,~W.~E. Rb and Cs Oligomers in Different Spin
	Configurations on Helium Nanodroplets. \emph{J. Phys. Chem. A} \textbf{2011},
	\emph{115}, 7005--7009\relax
	\mciteBstWouldAddEndPuncttrue
	\mciteSetBstMidEndSepPunct{\mcitedefaultmidpunct}
	{\mcitedefaultendpunct}{\mcitedefaultseppunct}\relax
	\EndOfBibitem
	\bibitem[Zhang and Drabbels(2012)Zhang, and Drabbels]{Zhang:2012}
	Zhang,~X.; Drabbels,~M. Communication: Barium ions and helium nanodroplets:
	Solvation and desolvation. \emph{J. Chem. Phys.} \textbf{2012}, \emph{137},
	051102\relax
	\mciteBstWouldAddEndPuncttrue
	\mciteSetBstMidEndSepPunct{\mcitedefaultmidpunct}
	{\mcitedefaultendpunct}{\mcitedefaultseppunct}\relax
	\EndOfBibitem
	\bibitem[Kautsch \latin{et~al.}(2012)Kautsch, Hasewend, Koch, and
	Ernst]{KautschPRA:2012}
	Kautsch,~A.; Hasewend,~M.; Koch,~M.; Ernst,~W.~E. Fano resonances in chromium
	photoionization spectra after photoinduced ejection from a superfluid helium
	nanodroplet. \emph{Phys. Rev. A} \textbf{2012}, \emph{86}, 033428\relax
	\mciteBstWouldAddEndPuncttrue
	\mciteSetBstMidEndSepPunct{\mcitedefaultmidpunct}
	{\mcitedefaultendpunct}{\mcitedefaultseppunct}\relax
	\EndOfBibitem
	\bibitem[Fechner \latin{et~al.}(2012)Fechner, Gr{\"u}ner, Sieg, Callegari,
	Ancilotto, Stienkemeier, and Mudrich]{Fechner:2012}
	Fechner,~L.; Gr{\"u}ner,~B.; Sieg,~A.; Callegari,~C.; Ancilotto,~F.;
	Stienkemeier,~F.; Mudrich,~M. Photoionization and imaging spectroscopy of
	rubidium atoms attached to helium nanodroplets. \emph{Phys. Chem. Chem.
		Phys.} \textbf{2012}, \emph{14}, 3843 –-- 3851\relax
	\mciteBstWouldAddEndPuncttrue
	\mciteSetBstMidEndSepPunct{\mcitedefaultmidpunct}
	{\mcitedefaultendpunct}{\mcitedefaultseppunct}\relax
	\EndOfBibitem
	\bibitem[von Vangerow \latin{et~al.}(2014)von Vangerow, Sieg, Stienkemeier,
	Mudrich, Leal, Mateo, Hernando, Barranco, and Pi]{Vangerow:2014}
	von Vangerow,~J.; Sieg,~A.; Stienkemeier,~F.; Mudrich,~M.; Leal,~A.; Mateo,~D.;
	Hernando,~A.; Barranco,~M.; Pi,~M. Desorption Dynamics of Heavy Alkali Metal
	Atoms (Rb, Cs) Off the Surface of Helium Nanodroplets. \emph{J. Phys. Chem.
		A} \textbf{2014}, \emph{118}, 6604--6614\relax
	\mciteBstWouldAddEndPuncttrue
	\mciteSetBstMidEndSepPunct{\mcitedefaultmidpunct}
	{\mcitedefaultendpunct}{\mcitedefaultseppunct}\relax
	\EndOfBibitem
	\bibitem[von Vangerow \latin{et~al.}(2015)von Vangerow, John, Stienkemeier, and
	Mudrich]{Vangerow:2015}
	von Vangerow,~J.; John,~O.; Stienkemeier,~F.; Mudrich,~M. Dynamics of solvation
	and desolvation of rubidium attached to He nanodroplets. \emph{J. Chem.
		Phys.} \textbf{2015}, \emph{143}, 034302\relax
	\mciteBstWouldAddEndPuncttrue
	\mciteSetBstMidEndSepPunct{\mcitedefaultmidpunct}
	{\mcitedefaultendpunct}{\mcitedefaultseppunct}\relax
	\EndOfBibitem
	\bibitem[Br{\"u}hl \latin{et~al.}(2001)Br{\"u}hl, Trasca, and
	Ernst]{Bruehl:2001}
	Br{\"u}hl,~F.~R.; Trasca,~R.~A.; Ernst,~W.~E. Rb--He exciplex formation on
	helium nanodroplets. \emph{J. Chem. Phys.} \textbf{2001}, \emph{115},
	10220--10224\relax
	\mciteBstWouldAddEndPuncttrue
	\mciteSetBstMidEndSepPunct{\mcitedefaultmidpunct}
	{\mcitedefaultendpunct}{\mcitedefaultseppunct}\relax
	\EndOfBibitem
	\bibitem[D\"{o}ppner \latin{et~al.}(2007)D\"{o}ppner, Diederich, G\"{o}de,
	Przystawik, Tiggesb\"{a}umker, and Meiwes-Broer]{DoeppnerJCP:2007}
	D\"{o}ppner,~T.; Diederich,~T.; G\"{o}de,~S.; Przystawik,~A.;
	Tiggesb\"{a}umker,~J.; Meiwes-Broer,~K.-H. Ion induced snowballs as a
	diagnostic tool to investigate the caging of metal clusters in large helium
	droplets. \emph{J. Chem. Phys.} \textbf{2007}, \emph{126}, 244513\relax
	\mciteBstWouldAddEndPuncttrue
	\mciteSetBstMidEndSepPunct{\mcitedefaultmidpunct}
	{\mcitedefaultendpunct}{\mcitedefaultseppunct}\relax
	\EndOfBibitem
	\bibitem[Mudrich \latin{et~al.}(2008)Mudrich, Droppelmann, Claas, Schulz, and
	Stienkemeier]{Mudrich:2008}
	Mudrich,~M.; Droppelmann,~G.; Claas,~P.; Schulz,~C.; Stienkemeier,~F. Quantum
	interference spectroscopy of RbHe exciplexes formed on helium nanodroplets.
	\emph{Phys. Rev. Lett.} \textbf{2008}, \emph{100}, 023401\relax
	\mciteBstWouldAddEndPuncttrue
	\mciteSetBstMidEndSepPunct{\mcitedefaultmidpunct}
	{\mcitedefaultendpunct}{\mcitedefaultseppunct}\relax
	\EndOfBibitem
	\bibitem[Loginov \latin{et~al.}(2011)Loginov, Callegari, Ancilotto, and
	Drabbels]{Loginov:2011}
	Loginov,~E.; Callegari,~C.; Ancilotto,~F.; Drabbels,~M. Spectroscopy on Rydberg
	States of Sodium Atoms on the Surface of Helium Nanodroplets. \emph{J. Phys.
		Chem. A} \textbf{2011}, \emph{115}, 6779--6788\relax
	\mciteBstWouldAddEndPuncttrue
	\mciteSetBstMidEndSepPunct{\mcitedefaultmidpunct}
	{\mcitedefaultendpunct}{\mcitedefaultseppunct}\relax
	\EndOfBibitem
	\bibitem[Giese \latin{et~al.}(2012)Giese, Mullins, Gr{\"u}ner, Weidem{\"u}ller,
	Stienkemeier, and Mudrich]{Giese:2012}
	Giese,~C.; Mullins,~T.; Gr{\"u}ner,~B.; Weidem{\"u}ller,~M.; Stienkemeier,~F.;
	Mudrich,~M. Formation and relaxation of RbHe exciplexes on He nanodroplets
	studied by femtosecond pump and picosecond probe spectroscopy. \emph{J. Chem.
		Phys.} \textbf{2012}, \emph{137}, 244307\relax
	\mciteBstWouldAddEndPuncttrue
	\mciteSetBstMidEndSepPunct{\mcitedefaultmidpunct}
	{\mcitedefaultendpunct}{\mcitedefaultseppunct}\relax
	\EndOfBibitem
	\bibitem[Leal \latin{et~al.}(2014)Leal, Mateo, Hernando, Pi, Barranco, Ponti,
	Cargnoni, and Drabbels]{Leal:2014}
	Leal,~A.; Mateo,~D.; Hernando,~A.; Pi,~M.; Barranco,~M.; Ponti,~A.;
	Cargnoni,~F.; Drabbels,~M. Picosecond solvation dynamics of alkali cations in
	superfluid $^{4}\mathrm{He}$ nanodroplets. \emph{Phys. Rev. B} \textbf{2014},
	\emph{90}, 224518\relax
	\mciteBstWouldAddEndPuncttrue
	\mciteSetBstMidEndSepPunct{\mcitedefaultmidpunct}
	{\mcitedefaultendpunct}{\mcitedefaultseppunct}\relax
	\EndOfBibitem
	\bibitem[Przystawik \latin{et~al.}(2008)Przystawik, G{\"o}de, D{\"o}ppner,
	Tiggesb{\"a}umker, and Meiwes-Broer]{Przystawik:2008}
	Przystawik,~A.; G{\"o}de,~S.; D{\"o}ppner,~T.; Tiggesb{\"a}umker,~J.;
	Meiwes-Broer,~K.-H. Light-induced collapse of metastable magnesium complexes
	formed in helium nanodroplets. \emph{Phys. Rev. A} \textbf{2008}, \emph{78},
	021202\relax
	\mciteBstWouldAddEndPuncttrue
	\mciteSetBstMidEndSepPunct{\mcitedefaultmidpunct}
	{\mcitedefaultendpunct}{\mcitedefaultseppunct}\relax
	\EndOfBibitem
	\bibitem[G{\"o}de \latin{et~al.}(2013)G{\"o}de, Irsig, Tiggesb{\"a}umker, and
	Meiwes-Broer]{Goede:2013}
	G{\"o}de,~S.; Irsig,~R.; Tiggesb{\"a}umker,~J.; Meiwes-Broer,~K.-H.
	Time-resolved studies on the collapse of magnesium atom foam in helium
	nanodroplets. \emph{New J. Phys.} \textbf{2013}, \emph{15}, 015026\relax
	\mciteBstWouldAddEndPuncttrue
	\mciteSetBstMidEndSepPunct{\mcitedefaultmidpunct}
	{\mcitedefaultendpunct}{\mcitedefaultseppunct}\relax
	\EndOfBibitem
	\bibitem[Pentlehner \latin{et~al.}(2013)Pentlehner, Nielsen, Slenczka,
	M\o{}lmer, and Stapelfeldt]{PentlehnerPRL:2013}
	Pentlehner,~D.; Nielsen,~J.~H.; Slenczka,~A.; M\o{}lmer,~K.; Stapelfeldt,~H.
	Impulsive Laser Induced Alignment of Molecules Dissolved in Helium
	Nanodroplets. \emph{Phys. Rev. Lett.} \textbf{2013}, \emph{110}, 093002\relax
	\mciteBstWouldAddEndPuncttrue
	\mciteSetBstMidEndSepPunct{\mcitedefaultmidpunct}
	{\mcitedefaultendpunct}{\mcitedefaultseppunct}\relax
	\EndOfBibitem
	\bibitem[Claas \latin{et~al.}(2006)Claas, Droppelmann, Schulz, Mudrich, and
	Stienkemeier]{Claas:2006}
	Claas,~P.; Droppelmann,~G.; Schulz,~C.~P.; Mudrich,~M.; Stienkemeier,~F. {Wave
		packet dynamics of potassium dimers attached to helium nanodroplets}.
	\emph{J. Phys. B} \textbf{2006}, \emph{39}, S1151 --– S1168\relax
	\mciteBstWouldAddEndPuncttrue
	\mciteSetBstMidEndSepPunct{\mcitedefaultmidpunct}
	{\mcitedefaultendpunct}{\mcitedefaultseppunct}\relax
	\EndOfBibitem
	\bibitem[Claas \latin{et~al.}(2007)Claas, Droppelmann, Schulz, Mudrich, and
	Stienkemeier]{Claas:2007}
	Claas,~P.; Droppelmann,~G.; Schulz,~C.~P.; Mudrich,~M.; Stienkemeier,~F. {Wave
		packet dynamics in the triplet states of Na$_2$ attached to helium
		nanodroplets}. \emph{J. Phys. Chem. A} \textbf{2007}, \emph{111}, 7537 –--
	7541\relax
	\mciteBstWouldAddEndPuncttrue
	\mciteSetBstMidEndSepPunct{\mcitedefaultmidpunct}
	{\mcitedefaultendpunct}{\mcitedefaultseppunct}\relax
	\EndOfBibitem
	\bibitem[Mudrich \latin{et~al.}(2009)Mudrich, Heister, Hippler, Giese, Dulieu,
	and Stienkemeier]{Mudrich:2009}
	Mudrich,~M.; Heister,~P.; Hippler,~T.; Giese,~C.; Dulieu,~O.; Stienkemeier,~F.
	Spectroscopy of triplet states of Rb$_2$ by femtosecond pump-probe
	photoionization of doped helium nanodroplets. \emph{Phys. Rev. A}
	\textbf{2009}, \emph{80}, 042512\relax
	\mciteBstWouldAddEndPuncttrue
	\mciteSetBstMidEndSepPunct{\mcitedefaultmidpunct}
	{\mcitedefaultendpunct}{\mcitedefaultseppunct}\relax
	\EndOfBibitem
	\bibitem[Gr{\"u}ner \latin{et~al.}(2011)Gr{\"u}ner, Schlesinger, Heister,
	Strunz, Stienkemeier, and Mudrich]{Gruner:2011}
	Gr{\"u}ner,~B.; Schlesinger,~M.; Heister,~P.; Strunz,~W.~T.; Stienkemeier,~F.;
	Mudrich,~M. Vibrational relaxation and decoherence of Rb$_2$ attached to
	helium nanodroplets. \emph{Phys. Chem. Chem. Phys.} \textbf{2011}, \emph{13},
	6816--6826\relax
	\mciteBstWouldAddEndPuncttrue
	\mciteSetBstMidEndSepPunct{\mcitedefaultmidpunct}
	{\mcitedefaultendpunct}{\mcitedefaultseppunct}\relax
	\EndOfBibitem
	\bibitem[Giese \latin{et~al.}(2011)Giese, Stienkemeier, Mudrich, Hauser, and
	Ernst]{Giese:2011}
	Giese,~C.; Stienkemeier,~F.; Mudrich,~M.; Hauser,~A.~W.; Ernst,~W.~E. Homo- and
	heteronuclear alkali metal trimers formed on helium nanodroplets. Part II.
	Femtosecond spectroscopy and spectra assignments. \emph{Phys. Chem. Chem.
		Phys.} \textbf{2011}, \emph{13}, 18769--18780\relax
	\mciteBstWouldAddEndPuncttrue
	\mciteSetBstMidEndSepPunct{\mcitedefaultmidpunct}
	{\mcitedefaultendpunct}{\mcitedefaultseppunct}\relax
	\EndOfBibitem
	\bibitem[Schlesinger \latin{et~al.}(2010)Schlesinger, Mudrich, Stienkemeier,
	and Strunz]{Schlesinger:2010}
	Schlesinger,~M.; Mudrich,~M.; Stienkemeier,~F.; Strunz,~W.~T. Dissipative
	vibrational wave packet dynamics of alkali dimers attached to helium
	nanodroplets. \emph{Chem. Phys. Lett.} \textbf{2010}, \emph{490},
	245--248\relax
	\mciteBstWouldAddEndPuncttrue
	\mciteSetBstMidEndSepPunct{\mcitedefaultmidpunct}
	{\mcitedefaultendpunct}{\mcitedefaultseppunct}\relax
	\EndOfBibitem
	\bibitem[Loginov and Drabbels(2014)Loginov, and Drabbels]{Loginov:2014}
	Loginov,~E.; Drabbels,~M. Dynamics of Excited Sodium Atoms Attached to Helium
	Nanodroplets. \emph{J. Phys. Chem. A} \textbf{2014}, \emph{118},
	2738--2748\relax
	\mciteBstWouldAddEndPuncttrue
	\mciteSetBstMidEndSepPunct{\mcitedefaultmidpunct}
	{\mcitedefaultendpunct}{\mcitedefaultseppunct}\relax
	\EndOfBibitem
	\bibitem[Loginov \latin{et~al.}(2015)Loginov, Hernando, Beswick, Halberstadt,
	and Drabbels]{Loginov:2015}
	Loginov,~E.; Hernando,~A.; Beswick,~J.~A.; Halberstadt,~N.; Drabbels,~M.
	Excitation of Sodium Atoms Attached to Helium Nanodroplets: The
	3p$\leftarrow$3s Transition Revisited. \emph{J. Phys. Chem. A} \textbf{2015},
	\emph{119}, 6033--6044\relax
	\mciteBstWouldAddEndPuncttrue
	\mciteSetBstMidEndSepPunct{\mcitedefaultmidpunct}
	{\mcitedefaultendpunct}{\mcitedefaultseppunct}\relax
	\EndOfBibitem
	\bibitem[Dalfovo(1994)]{Dalfovo:1994}
	Dalfovo,~F. Atomic and molecular impurities in $^4${H}e clusters. \emph{Z.
		Phys. D} \textbf{1994}, \emph{29}, 61--66\relax
	\mciteBstWouldAddEndPuncttrue
	\mciteSetBstMidEndSepPunct{\mcitedefaultmidpunct}
	{\mcitedefaultendpunct}{\mcitedefaultseppunct}\relax
	\EndOfBibitem
	\bibitem[Ancilotto \latin{et~al.}(1995)Ancilotto, DeToffol, and
	Toigo]{Ancilotto:1995}
	Ancilotto,~F.; DeToffol,~G.; Toigo,~F. Sodium dimers on the surface of liquid
	$^4${H}e. \emph{Phys. Rev. B} \textbf{1995}, \emph{52}, 16125--16129\relax
	\mciteBstWouldAddEndPuncttrue
	\mciteSetBstMidEndSepPunct{\mcitedefaultmidpunct}
	{\mcitedefaultendpunct}{\mcitedefaultseppunct}\relax
	\EndOfBibitem
	\bibitem[Stienkemeier \latin{et~al.}(1995)Stienkemeier, Ernst, Higgins, and
	Scoles]{Stienkemeier:1995}
	Stienkemeier,~F.; Ernst,~W.~E.; Higgins,~J.; Scoles,~G. On the use of liquid
	helium cluster beams for the preparation and spectroscopy of the triplet
	states of alkli dimers and other weakly bound complexes. \emph{J. Chem.
		Phys.} \textbf{1995}, \emph{102}, 615--617\relax
	\mciteBstWouldAddEndPuncttrue
	\mciteSetBstMidEndSepPunct{\mcitedefaultmidpunct}
	{\mcitedefaultendpunct}{\mcitedefaultseppunct}\relax
	\EndOfBibitem
	\bibitem[Reho \latin{et~al.}(2000)Reho, Higgins, Callegari, Lehmann, and
	Scoles]{Reho:2000}
	Reho,~J.; Higgins,~J.; Callegari,~C.; Lehmann,~K.~K.; Scoles,~G. Alkali-helium
	exciplex formation on the surface of helium nanodroplets. {I}. {D}ispersed
	emission spectroscopy. \emph{J. Chem. Phys.} \textbf{2000}, \emph{113},
	9686--9693\relax
	\mciteBstWouldAddEndPuncttrue
	\mciteSetBstMidEndSepPunct{\mcitedefaultmidpunct}
	{\mcitedefaultendpunct}{\mcitedefaultseppunct}\relax
	\EndOfBibitem
	\bibitem[Schulz \latin{et~al.}(2001)Schulz, Claas, and
	Stienkemeier]{Schulz:2001}
	Schulz,~C.~P.; Claas,~P.; Stienkemeier,~F. Formation of {K}$^*${He} exciplexes
	on the surface of helium nanodroplets studied in real time. \emph{Phys. Rev.
		Lett.} \textbf{2001}, \emph{87}, 153401\relax
	\mciteBstWouldAddEndPuncttrue
	\mciteSetBstMidEndSepPunct{\mcitedefaultmidpunct}
	{\mcitedefaultendpunct}{\mcitedefaultseppunct}\relax
	\EndOfBibitem
	\bibitem[Callegari and Ancilotto(2011)Callegari, and Ancilotto]{Callegari:2011}
	Callegari,~C.; Ancilotto,~F. Perturbation Method to Calculate the Interaction
	Potentials and Electronic Excitation Spectra of Atoms in He Nanodroplets.
	\emph{J. Phys. Chem. A} \textbf{2011}, \emph{115}, 6789--6796\relax
	\mciteBstWouldAddEndPuncttrue
	\mciteSetBstMidEndSepPunct{\mcitedefaultmidpunct}
	{\mcitedefaultendpunct}{\mcitedefaultseppunct}\relax
	\EndOfBibitem
	\bibitem[Aub{\"o}ck \latin{et~al.}(2008)Aub{\"o}ck, Nagl, Callegari, and
	Ernst]{Auboeck:2008}
	Aub{\"o}ck,~G.; Nagl,~J.; Callegari,~C.; Ernst,~W.~E. Electron Spin Pumping of
	Rb Atoms on He Nanodroplets via Nondestructive Optical Excitation.
	\emph{Phys. Rev. Lett.} \textbf{2008}, \emph{101}, 035301\relax
	\mciteBstWouldAddEndPuncttrue
	\mciteSetBstMidEndSepPunct{\mcitedefaultmidpunct}
	{\mcitedefaultendpunct}{\mcitedefaultseppunct}\relax
	\EndOfBibitem
	\bibitem[Buzzacchi \latin{et~al.}(2001)Buzzacchi, Galli, and
	Reatto]{Buzzacchi:2001}
	Buzzacchi,~M.; Galli,~D.~E.; Reatto,~L. Alkali ions in superfluid $^4$He and
	structure of the snowball. \emph{Phys. Rev. B} \textbf{2001}, \emph{64},
	094512\relax
	\mciteBstWouldAddEndPuncttrue
	\mciteSetBstMidEndSepPunct{\mcitedefaultmidpunct}
	{\mcitedefaultendpunct}{\mcitedefaultseppunct}\relax
	\EndOfBibitem
	\bibitem[M\"{u}ller \latin{et~al.}(2009)M\"{u}ller, Mudrich, and
	Stienkemeier]{Mueller:2009}
	M\"{u}ller,~S.; Mudrich,~M.; Stienkemeier,~F. Alkali-helium snowball complexes
	formed on helium nanodroplets. \emph{J. Chem. Phys.} \textbf{2009},
	\emph{131}, 044319\relax
	\mciteBstWouldAddEndPuncttrue
	\mciteSetBstMidEndSepPunct{\mcitedefaultmidpunct}
	{\mcitedefaultendpunct}{\mcitedefaultseppunct}\relax
	\EndOfBibitem
	\bibitem[Theisen \latin{et~al.}(2010)Theisen, Lackner, and Ernst]{Theisen:2010}
	Theisen,~M.; Lackner,~F.; Ernst,~W.~E. Forming Rb$^+$ snowballs in the center
	of He nanodroplets. \emph{Phys. Chem. Chem. Phys.} \textbf{2010}, \emph{12},
	14861--14863\relax
	\mciteBstWouldAddEndPuncttrue
	\mciteSetBstMidEndSepPunct{\mcitedefaultmidpunct}
	{\mcitedefaultendpunct}{\mcitedefaultseppunct}\relax
	\EndOfBibitem
	\bibitem[Stienkemeier \latin{et~al.}(1996)Stienkemeier, Higgins, Callegari,
	Kanorsky, Ernst, and Scoles]{Stienkemeier:1996}
	Stienkemeier,~F.; Higgins,~J.; Callegari,~C.; Kanorsky,~S.~I.; Ernst,~W.~E.;
	Scoles,~G. Spectroscopy of alkali atoms ({Li, Na, K}) attached to large
	helium clusters. \emph{Z. Phys. D} \textbf{1996}, \emph{38}, 253--263\relax
	\mciteBstWouldAddEndPuncttrue
	\mciteSetBstMidEndSepPunct{\mcitedefaultmidpunct}
	{\mcitedefaultendpunct}{\mcitedefaultseppunct}\relax
	\EndOfBibitem
	\bibitem[Loginov and Drabbels(2011)Loginov, and Drabbels]{LoginovPRL:2011}
	Loginov,~E.; Drabbels,~M. Unusual Rydberg System Consisting of a Positively
	Charged Helium Nanodroplet with an Orbiting Electron. \emph{Phys. Rev. Lett.}
	\textbf{2011}, \emph{106}, 083401\relax
	\mciteBstWouldAddEndPuncttrue
	\mciteSetBstMidEndSepPunct{\mcitedefaultmidpunct}
	{\mcitedefaultendpunct}{\mcitedefaultseppunct}\relax
	\EndOfBibitem
	\bibitem[B{\"u}nermann \latin{et~al.}(2007)B{\"u}nermann, Droppelmann,
	Hernando, Mayol, and Stienkemeier]{Buenermann:2007}
	B{\"u}nermann,~O.; Droppelmann,~G.; Hernando,~A.; Mayol,~R.; Stienkemeier,~F.
	Unraveling the Absorption Spectra of Alkali Metal Atoms Attached to Helium
	Nanodroplets. \emph{J. Phys. Chem. A} \textbf{2007}, \emph{111}, 12684 --
	12694\relax
	\mciteBstWouldAddEndPuncttrue
	\mciteSetBstMidEndSepPunct{\mcitedefaultmidpunct}
	{\mcitedefaultendpunct}{\mcitedefaultseppunct}\relax
	\EndOfBibitem
	\bibitem[Lackner \latin{et~al.}(2011)Lackner, Krois, Theisen, Koch, and
	Ernst]{Lackner:2011}
	Lackner,~F.; Krois,~G.; Theisen,~M.; Koch,~M.; Ernst,~W.~E. Spectroscopy of
	nS{,} nP{,} and nD Rydberg series of Cs atoms on helium nanodroplets.
	\emph{Phys. Chem. Chem. Phys.} \textbf{2011}, \emph{13}, 18781--18788\relax
	\mciteBstWouldAddEndPuncttrue
	\mciteSetBstMidEndSepPunct{\mcitedefaultmidpunct}
	{\mcitedefaultendpunct}{\mcitedefaultseppunct}\relax
	\EndOfBibitem
	\bibitem[Stark and Kresin(2010)Stark, and Kresin]{Stark:2010}
	Stark,~C.; Kresin,~V.~V. Critical sizes for the submersion of alkali clusters
	into liquid helium. \emph{Phys. Rev. B} \textbf{2010}, \emph{81},
	085401\relax
	\mciteBstWouldAddEndPuncttrue
	\mciteSetBstMidEndSepPunct{\mcitedefaultmidpunct}
	{\mcitedefaultendpunct}{\mcitedefaultseppunct}\relax
	\EndOfBibitem
	\bibitem[Stienkemeier and Vilesov(2001)Stienkemeier, and
	Vilesov]{Stienkemeier:2001}
	Stienkemeier,~F.; Vilesov,~A.~F. Electronic spectroscopy in {H}e droplets.
	\emph{J. Chem. Phys.} \textbf{2001}, \emph{115}, 10119--10137\relax
	\mciteBstWouldAddEndPuncttrue
	\mciteSetBstMidEndSepPunct{\mcitedefaultmidpunct}
	{\mcitedefaultendpunct}{\mcitedefaultseppunct}\relax
	\EndOfBibitem
	\bibitem[Toennies and Vilesov(2004)Toennies, and Vilesov]{Toennies:2004}
	Toennies,~J.~P.; Vilesov,~A.~F. Superfluid helium droplets: A uniquely cold
	nanomatrix for molecules and molecular complexes. \emph{Angew. Chem. Int.
		Ed.} \textbf{2004}, \emph{43}, 2622--2648\relax
	\mciteBstWouldAddEndPuncttrue
	\mciteSetBstMidEndSepPunct{\mcitedefaultmidpunct}
	{\mcitedefaultendpunct}{\mcitedefaultseppunct}\relax
	\EndOfBibitem
	\bibitem[Stienkemeier \latin{et~al.}(1995)Stienkemeier, Higgins, Ernst, and
	Scoles]{Stienkemeier2:1995}
	Stienkemeier,~F.; Higgins,~J.; Ernst,~W.~E.; Scoles,~G. Laser spectroscopy of
	alkali-doped helium clusters. \emph{Phys. Rev. Lett.} \textbf{1995},
	\emph{74}, 3592--3595\relax
	\mciteBstWouldAddEndPuncttrue
	\mciteSetBstMidEndSepPunct{\mcitedefaultmidpunct}
	{\mcitedefaultendpunct}{\mcitedefaultseppunct}\relax
	\EndOfBibitem
	\bibitem[Nagl \latin{et~al.}(2008)Nagl, Aub{\"o}ck, Hauser, Allard, Callegari,
	and Ernst]{Nagl_jcp:2008}
	Nagl,~J.; Aub{\"o}ck,~G.; Hauser,~A.; Allard,~O.; Callegari,~C.; Ernst,~W.
	High-Spin Alkali Trimers on Helium Nanodroplets: Spectral separation and
	analysis. \emph{J. Chem. Phys.} \textbf{2008}, \emph{128}, 154320\relax
	\mciteBstWouldAddEndPuncttrue
	\mciteSetBstMidEndSepPunct{\mcitedefaultmidpunct}
	{\mcitedefaultendpunct}{\mcitedefaultseppunct}\relax
	\EndOfBibitem
	\bibitem[Garcia \latin{et~al.}(2004)Garcia, Nahon, and Powis]{Garcia:2004}
	Garcia,~G.~A.; Nahon,~L.; Powis,~I. Two-dimensional charged particle image
	inversion using a polar basis function expansion. \emph{Rev. Sci. Instrum.}
	\textbf{2004}, \emph{75}, 4989--4996\relax
	\mciteBstWouldAddEndPuncttrue
	\mciteSetBstMidEndSepPunct{\mcitedefaultmidpunct}
	{\mcitedefaultendpunct}{\mcitedefaultseppunct}\relax
	\EndOfBibitem
	\bibitem[Lozeille \latin{et~al.}(2006)Lozeille, Fioretti, Gabbanini, Huang,
	Pechkis, Wang, Gould, Eyler, Stwalley, Aymar, and Dulieu]{Lozeille:2006}
	Lozeille,~J.; Fioretti,~A.; Gabbanini,~C.; Huang,~Y.; Pechkis,~H.; Wang,~D.;
	Gould,~P.; Eyler,~E.; Stwalley,~W.; Aymar,~M.; Dulieu,~O. Detection by
	two-photon ionization and magnetic trapping of cold Rb$_2$ triplet state
	molecules. \emph{Eur. Phys. J. D} \textbf{2006}, \emph{39}, 261 -- 269\relax
	\mciteBstWouldAddEndPuncttrue
	\mciteSetBstMidEndSepPunct{\mcitedefaultmidpunct}
	{\mcitedefaultendpunct}{\mcitedefaultseppunct}\relax
	\EndOfBibitem
	\bibitem[Lee \latin{et~al.}(2000)Lee, Yoon, Baek, Joo, seok Ryu, and
	Kim]{Lee:2000}
	Lee,~Y.; Yoon,~Y.; Baek,~S.~J.; Joo,~D.-L.; seok Ryu,~J.; Kim,~B. Direct
	observation of the $2^3\Pi_u$ state of Rb$_2$ in a pulsed molecular beam:
	Rotational branch intensity anomalies in the $2^3\Pi_u(1_u) -
	X^1\Sigma^+_g(0^+_g)$ bands. \emph{J. Chem. Phys.} \textbf{2000}, \emph{113},
	2116--2123\relax
	\mciteBstWouldAddEndPuncttrue
	\mciteSetBstMidEndSepPunct{\mcitedefaultmidpunct}
	{\mcitedefaultendpunct}{\mcitedefaultseppunct}\relax
	\EndOfBibitem
	\bibitem[Mudrich \latin{et~al.}(2004)Mudrich, B{\"u}nermann, Stienkemeier,
	Dulieu, and Weidem{\"u}ller]{Mudrich:2004}
	Mudrich,~M.; B{\"u}nermann,~O.; Stienkemeier,~F.; Dulieu,~O.;
	Weidem{\"u}ller,~M. Formation of cold bialkali dimers on helium nanodroplets.
	\emph{Eur. Phys. J. D} \textbf{2004}, \emph{31}, 291--299\relax
	\mciteBstWouldAddEndPuncttrue
	\mciteSetBstMidEndSepPunct{\mcitedefaultmidpunct}
	{\mcitedefaultendpunct}{\mcitedefaultseppunct}\relax
	\EndOfBibitem
	\bibitem[Allard \latin{et~al.}(2006)Allard, Nagl, Aub{\"o}ck, Callegari, and
	Ernst]{Allard:2006}
	Allard,~O.; Nagl,~J.; Aub{\"o}ck,~G.; Callegari,~C.; Ernst,~W.~E. Investigation
	of KRb and Rb$_2$ formed on cold helium nanodroplets. \emph{J. Phys. B}
	\textbf{2006}, \emph{39}, S1169 --– S1181\relax
	\mciteBstWouldAddEndPuncttrue
	\mciteSetBstMidEndSepPunct{\mcitedefaultmidpunct}
	{\mcitedefaultendpunct}{\mcitedefaultseppunct}\relax
	\EndOfBibitem
	\bibitem[Aub{\"o}ck \latin{et~al.}(2010)Aub{\"o}ck, Aymar, Dulieu, and
	Ernst]{Auboeck:2010}
	Aub{\"o}ck,~G.; Aymar,~M.; Dulieu,~O.; Ernst,~W.~E. Reinvestigation of the
	Rb$_2$ band on helium nanodroplets. \emph{J. Chem. Phys.} \textbf{2010},
	\emph{132}, 054304\relax
	\mciteBstWouldAddEndPuncttrue
	\mciteSetBstMidEndSepPunct{\mcitedefaultmidpunct}
	{\mcitedefaultendpunct}{\mcitedefaultseppunct}\relax
	\EndOfBibitem
	\bibitem[Loginov(2008)]{LoginovPhD:2008}
	Loginov,~E. {P}hotoexcitation and {P}hotoionization {D}ynamics of {D}oped
	{L}iquid {H}elium-4 {H}anodroplets. Ph.D.\ thesis, {\'E}cole {P}olytechnique
	{F}{\'e}d{\'e}rale de {L}ausanne, 2008\relax
	\mciteBstWouldAddEndPuncttrue
	\mciteSetBstMidEndSepPunct{\mcitedefaultmidpunct}
	{\mcitedefaultendpunct}{\mcitedefaultseppunct}\relax
	\EndOfBibitem
	\bibitem[Aub{\"o}ck \latin{et~al.}(2007)Aub{\"o}ck, Nagl, Callegari, and
	Ernst]{Auboeck:2007}
	Aub{\"o}ck,~G.; Nagl,~J.; Callegari,~C.; Ernst,~W. Triplet State Excitation of
	Alkali Molecules on Helium Droplets: Experiments and Theory. \emph{J. Phys.
		Chem. A} \textbf{2007}, \emph{111}, 7404 –-- 7410\relax
	\mciteBstWouldAddEndPuncttrue
	\mciteSetBstMidEndSepPunct{\mcitedefaultmidpunct}
	{\mcitedefaultendpunct}{\mcitedefaultseppunct}\relax
	\EndOfBibitem
	\bibitem[Ernst \latin{et~al.}(2006)Ernst, Huber, Jiang, Beuc, Movre, and
	Pichler]{Ernst:2006}
	Ernst,~W.~E.; Huber,~R.; Jiang,~S.; Beuc,~R.; Movre,~M.; Pichler,~G. Cesium
	dimer spectroscopy on helium droplets. \emph{J. Chem. Phys.} \textbf{2006},
	\emph{124}, 024313\relax
	\mciteBstWouldAddEndPuncttrue
	\mciteSetBstMidEndSepPunct{\mcitedefaultmidpunct}
	{\mcitedefaultendpunct}{\mcitedefaultseppunct}\relax
	\EndOfBibitem
	\bibitem[Park \latin{et~al.}(2001)Park, Suh, Lee, and Jeung]{Park:2001}
	Park,~S.~J.; Suh,~S.~W.; Lee,~Y.~S.; Jeung,~G.-H. Theoretical Study of the
	Electronic States of the {R}b$_2$ Molecule. \emph{J. Mol. Spectrosc.}
	\textbf{2001}, \emph{207}, 129--135\relax
	\mciteBstWouldAddEndPuncttrue
	\mciteSetBstMidEndSepPunct{\mcitedefaultmidpunct}
	{\mcitedefaultendpunct}{\mcitedefaultseppunct}\relax
	\EndOfBibitem
	\bibitem[LeRoy(1995)]{level}
	LeRoy,~R. Chemical Physics Research Report. University of Waterloo, CP-555,
	1995, 1995\relax
	\mciteBstWouldAddEndPuncttrue
	\mciteSetBstMidEndSepPunct{\mcitedefaultmidpunct}
	{\mcitedefaultendpunct}{\mcitedefaultseppunct}\relax
	\EndOfBibitem
	\bibitem[Zare(1972)]{Zare:1972}
	Zare,~R.~N. Photoejection Dynamics [1]. \emph{Mol. Photochem.} \textbf{1972},
	\emph{44}, 1 -- 37\relax
	\mciteBstWouldAddEndPuncttrue
	\mciteSetBstMidEndSepPunct{\mcitedefaultmidpunct}
	{\mcitedefaultendpunct}{\mcitedefaultseppunct}\relax
	\EndOfBibitem
	\bibitem[Mudholkar and Hutson(1999)Mudholkar, and Hutson]{Mudholkar:1999}
	Mudholkar,~G.~S.; Hutson,~A.~D. The epsilon-skew-normal distribution for
	analyzing near-normal data. \emph{J. Statist. Plann. Inference}
	\textbf{1999}, \emph{83}, 291--309\relax
	\mciteBstWouldAddEndPuncttrue
	\mciteSetBstMidEndSepPunct{\mcitedefaultmidpunct}
	{\mcitedefaultendpunct}{\mcitedefaultseppunct}\relax
	\EndOfBibitem
	\bibitem[Bovino \latin{et~al.}(2009)Bovino, Coccia, Bodo, Lopez-Durán, and
	Gianturco]{Bovino:2009}
	Bovino,~S.; Coccia,~E.; Bodo,~E.; Lopez-Durán,~D.; Gianturco,~F.~A.
	Spin-driven structural effects in alkali doped $^4$He clusters from quantum
	calculations. \emph{J. Chem. Phys.} \textbf{2009}, \emph{130}, 224903\relax
	\mciteBstWouldAddEndPuncttrue
	\mciteSetBstMidEndSepPunct{\mcitedefaultmidpunct}
	{\mcitedefaultendpunct}{\mcitedefaultseppunct}\relax
	\EndOfBibitem
	\bibitem[Leino \latin{et~al.}(2011)Leino, Viel, and Zillich]{Leino:2011}
	Leino,~M.; Viel,~A.; Zillich,~R.~E. Electronically excited rubidium atom in
	helium clusters and films. II. Second excited state and absorption spectrum.
	\emph{J. Chem. Phys.} \textbf{2011}, \emph{134}, 024316\relax
	\mciteBstWouldAddEndPuncttrue
	\mciteSetBstMidEndSepPunct{\mcitedefaultmidpunct}
	{\mcitedefaultendpunct}{\mcitedefaultseppunct}\relax
	\EndOfBibitem
	\bibitem[Schulz \latin{et~al.}(2004)Schulz, Claas, Schumacher, and
	Stienkemeier]{Schulz:2004}
	Schulz,~C.~P.; Claas,~P.; Schumacher,~D.; Stienkemeier,~F. Formation and
	Stability of High-Spin Alkali Clusters. \emph{Phys. Rev. Lett.}
	\textbf{2004}, \emph{92}, 013401\relax
	\mciteBstWouldAddEndPuncttrue
	\mciteSetBstMidEndSepPunct{\mcitedefaultmidpunct}
	{\mcitedefaultendpunct}{\mcitedefaultseppunct}\relax
	\EndOfBibitem
	\bibitem[Droppelmann \latin{et~al.}(2009)Droppelmann, Mudrich, Schulz, and
	Stienkemeier]{Droppelmann:2009}
	Droppelmann,~G.; Mudrich,~M.; Schulz,~C.~P.; Stienkemeier,~F. Stability of
	two-component alkali clusters formed on helium nanodroplets. \emph{Eur. Phys.
		J. D} \textbf{2009}, \emph{52}, 67--70\relax
	\mciteBstWouldAddEndPuncttrue
	\mciteSetBstMidEndSepPunct{\mcitedefaultmidpunct}
	{\mcitedefaultendpunct}{\mcitedefaultseppunct}\relax
	\EndOfBibitem
\end{mcitethebibliography}

\providecommand{\latin}[1]{#1}
\providecommand*\mcitethebibliography{\thebibliography}
\csname @ifundefined\endcsname{endmcitethebibliography}
{\let\endmcitethebibliography\endthebibliography}{}

\end{document}